%% file: missing_data_manuscript_final.tex
\documentclass[conference]{IEEEtran}
\usepackage{algorithm,algorithmic}
\usepackage[noadjust]{cite}
\usepackage{amssymb,mathrsfs, mathtools, amsthm}

\DeclarePairedDelimiter\floor{\lfloor}{\rfloor}
\DeclarePairedDelimiter{\norm}{\lVert}{\rVert}
\newtheorem{theorem}{Theorem}
\newtheorem{lemma}{Lemma}
\newtheorem{assumption}{Assumption}
\newtheorem{remark}{Remark}
\newtheorem{example}{Example}
\newtheorem{corollary}{Corollary}
\newtheorem{definition}{Definition}
\usepackage{xcolor}
\usepackage{tikz}
\usetikzlibrary{matrix,decorations.pathreplacing,calc}
\IEEEoverridecommandlockouts

\makeatletter
\let\NAT@parse\undefined
\newcommand{\removelatexerror}{\let\@latex@error\@gobble}
\makeatother
\usepackage{hyperref}  

\newcommand\copyrighttext{%
	\footnotesize \copyright 2024 IEEE. Personal use of this material is permitted. Permission from IEEE must be obtained for all other uses, in any current or future media, including reprinting/republishing this material for advertising or promotional purposes, creating new collective works, for resale or redistribution to servers or lists, or reuse of any copyrighted component of this work in other works.}
\newcommand\copyrightnotice{%
	\begin{tikzpicture}[remember picture,overlay]
		\node[anchor=south,yshift=5pt] at (current page.south) {\fbox{\parbox{\dimexpr\textwidth-\fboxsep-\fboxrule\relax}{\copyrighttext}}};
	\end{tikzpicture}%
}

\begin{document}
\title{\LARGE{\textbf{Data-based system representations from irregularly measured data}}}

\author{Mohammad Alsalti$^{1}$, Ivan Markovsky$^{2}$, Victor G. Lopez$^{1}$, and Matthias A. Müller$^{1}$ %
	\thanks{$^{1}$Leibniz University Hannover, Institute of Automatic Control, 30167 Hannover, Germany. E-mail:\{\href{maitlo:alsalti@irt.uni-hannover.de}{alsalti},\href{maitlo:lopez@irt.uni-hannover.de}{lopez},\href{maitlo:mueller@irt.uni-hannover.de}{mueller}\}@irt.uni-hannover.de}%
	\thanks{$^{2}$International Centre for Numerical Methods in Engineering (CIMNE), Gran Capitàn, Barcelona, Spain and the Catalan	Institution for Research and Advanced Studies (ICREA), Barcelona, Spain. Email: \href{maitlo:imarkovsky@cimne.upc.edu}{imarkovsky}@cimne.upc.edu}%
	\thanks{This work has received funding from the European Research Council (ERC) under the European Union’s Horizon 2020 research and innovation programme (grant agreement No 948679), as well as from the Catalan Institution for Research and Advanced Studies (ICREA) and the Fond for Scientific Research Vlaanderen (FWO) project G033822N.
	}
}

\maketitle%
\thispagestyle{empty}%
\copyrightnotice%

	\begin{abstract}
		Non-parametric representations of dynamical systems based on the image of a Hankel matrix of data are extensively used for data-driven control. However, if samples of data are missing, obtaining such representations becomes a difficult task. By exploiting the kernel structure of Hankel matrices of irregularly measured data generated by a linear time-invariant system, we provide computational methods for which any \textit{complete} finite-length behavior of the system can be obtained. For the special case of periodically missing outputs, we provide conditions on the input such that the former result is guaranteed. In the presence of noise in the data, our method returns an approximate finite-length behavior of the system. We illustrate our result with several examples, including its use for approximate data completion in real-world applications and compare it to alternative methods.

	\end{abstract}
	\IEEEpeerreviewmaketitle
	
	\input{sections/introduction}
	\input{sections/preliminaries}
	\input{sections/problemformulation}
	\input{sections/kernelstructure}
	\input{sections/periodic_SO}
	\input{sections/noisydata}
	\input{sections/examples}
	\input{sections/conclusion}
	
	\bibliographystyle{IEEEtran}
	\bibliography{references}
	
	\input{sections/appendix}
	
	\begin{IEEEbiography}[{\includegraphics[width=1in,height=1.25in,clip,keepaspectratio]{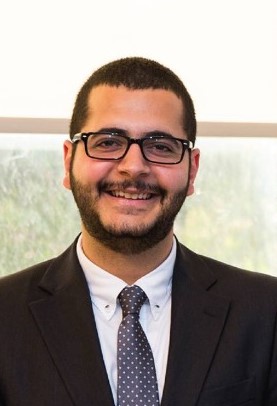}}]{Mohammad Alsalti}
		received his B.Sc. in Mechanical Engineering from the University of Jordan, Jordan, in 2017. In 2017, he was an intern at NASA Ames Research Center as part of the intelligent robotics group. In 2018, Mohammad was awarded the Fulbright scholarship to pursue graduate studies. In 2020, he obtained his M.Sc. in Mechanical Engineering from the University of Maryland, College Park, USA. He is currently a Ph.D. student at the Institute of Automatic Control at Leibniz University Hannover, Germany. He is working on developing data-driven control techniques for linear and nonlinear systems.
	\end{IEEEbiography}

	\begin{IEEEbiography}[{\includegraphics[width=1in,height=1.25in,clip,keepaspectratio]{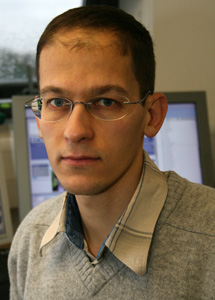}}]{Ivan Markovsky}
		received the Ph.D. degree in electrical engineering from the Katholieke Universiteit Leuven, Leuven, Belgium, in February 2005. He is currently an ICREA Professor with the International Centre for Numerical Methods in Engineering, Barcelona. From 2006 to 2012, he was an Assistant Professor with the School of Electronics and Computer Science, University of Southampton, Southampton, U.K., and from 2012 to 2022, an Associate Professor with the Vrije Universiteit, Brussel, Belgium. His research interests are computational methods for system theory, identification, and data-driven control in the behavioral setting. Dr. Markovsky was the recipient of an ERC starting grant “Structured low-rank approximation: Theory, algorithms, and applications” 2010–2015, Householder Prize honorable mention 2008, and research mandate by the Vrije Universiteit Brussel research council 2012–2022.
	\end{IEEEbiography}
	
	\begin{IEEEbiography}[{\includegraphics[width=1in,height=1.25in,clip,keepaspectratio]{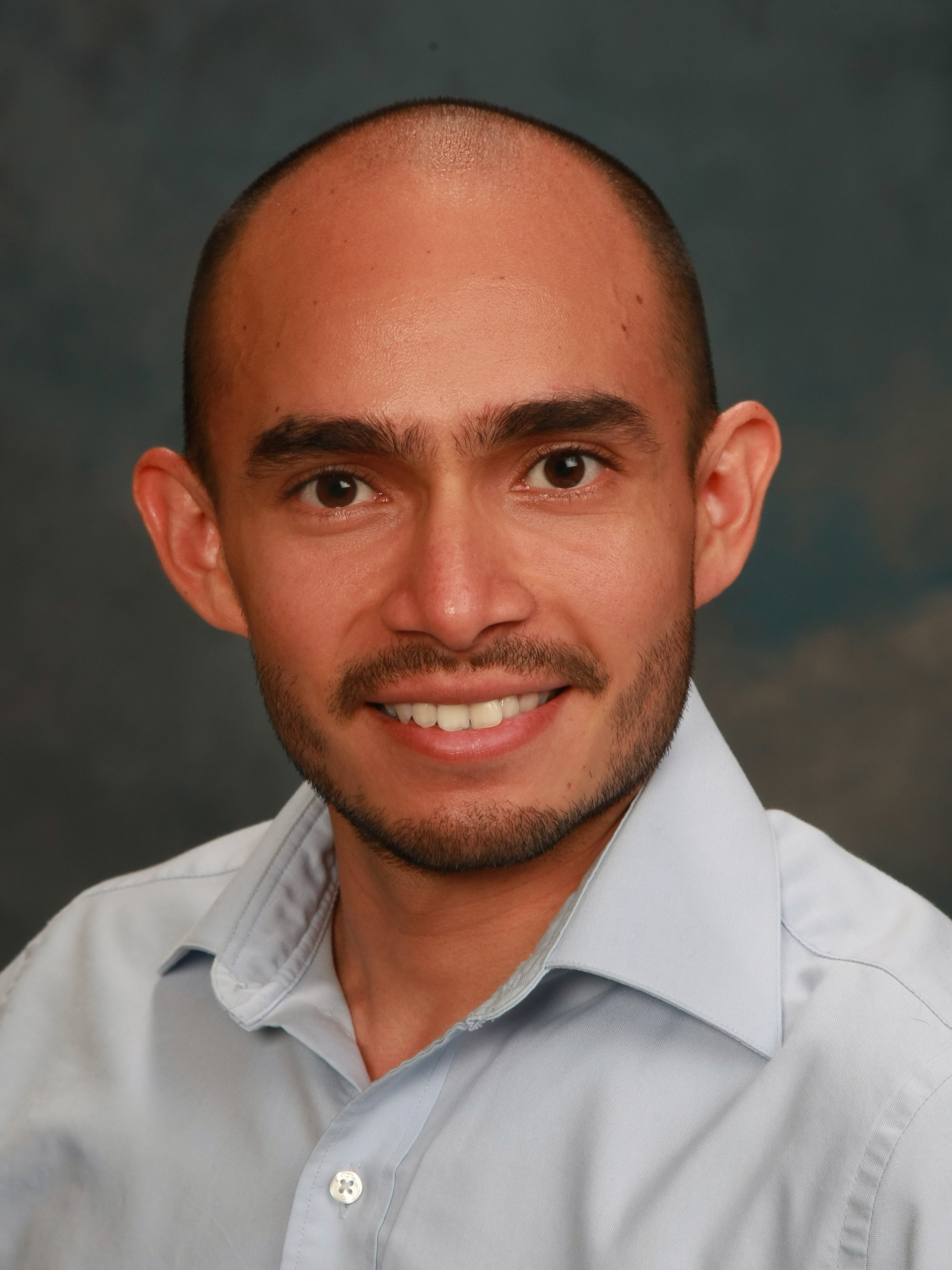}}]{Victor G. Lopez}
		received his B.Sc. degree in Communications and Electronics Engineering from the Universidad Autonoma de Campeche, in Campeche, Mexico, in 2010, the M.Sc. degree in Electrical Engineering from the Research and Advanced Studies Center (Cinvestav), in Guadalajara, Mexico, in 2013, and his Ph.D. degree in Electrical Engineering from the University of Texas at Arlington, Texas, USA, in 2019. In 2015 Victor was a Lecturer at the Western Technological Institute of Superior Studies (ITESO) in Guadalajara, Mexico. From August 2019 to June 2020, he was a postdoctoral researcher at the University of Texas at Arlington Research Institute and an Adjunct Professor in the Electrical Engineering department at UTA. Victor is currently a postdoctoral researcher at the Institute of Automatic Control, Leibniz University Hannover, in Hannover, Germany. His research interest include cyber-physical systems, reinforcement learning, game theory, distributed control and robust control. 
	\end{IEEEbiography}

	\begin{IEEEbiography}[{\includegraphics[width=1in,height=1.25in,clip,keepaspectratio]{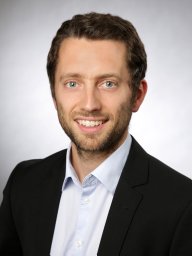}}]{Matthias A. Müller}
		Matthias A. Müller received a Diploma degree in Engineering Cybernetics from the University of Stuttgart, Germany, an M.Sc. in Electrical and Computer Engineering from the University of Illinois at Urbana-Champaign, US (both in 2009), and a Ph.D. from the University of Stuttgart in 2014. Since 2019, he has been director of the Institute of Automatic Control and full professor at the Leibniz University Hannover, Germany. His research interests include nonlinear control and estimation, model predictive control, and data- and learning-based control, with applications in different fields including biomedical engineering and robotics. He has received various awards for his work, including the 2015 European Systems \& Control PhD Thesis Award, the inaugural Brockett-Willems Outstanding Paper Award for the best paper published in Systems \& Control Letters in the period 2014-2018, an ERC starting grant in 2020, the IEEE CSS George S. Axelby Outstanding Paper Award 2022, and the Journal of Process Control Paper Award 2023. He serves as associate editor for Automatica and editor of the Intl. Journal of Robust and Nonlinear Control.
	\end{IEEEbiography}
\end{document}

%% file: sections/introduction.tex
\section{Introduction}\label{sec_introduction}
Many systems of interest are too complicated to model using first principles. However, abundance of data generated from such systems and recent advances in machine learning open new research possibilities for modeling and control design. Data- and learning-based models of dynamical systems can be obtained using, e.g., Neural networks \cite{Miller95}, Gaussian processes \cite{Beckers16}, or other machine learning techniques. Alternatively, the behavioral approach to dynamical systems theory \cite{Willems86} exploits system properties to arrive at non-parametric representations of dynamical systems.

A recently revived and celebrated result from the behavioral framework is the so-called fundamental lemma \cite{Willems05}. This lemma provides a non-parametric representation of the finite-length behavior of a discrete-time LTI system as the image of a Hankel matrix of a single persistently exciting (PE) trajectory. In particular, it asserts that any trajectory of an LTI system is given by a linear combination of time-shifts of a single measured trajectory. This result is central to several works on direct data-based control for linear \cite{Persis20, vanWaarde20, Berberich20, Pan22} and nonlinear systems \cite{Verhoek21, Escobedo20, Alsalti2022}. The reader is referred to \cite{Markovsky21} for a comprehensive review. Such direct data-based control approaches based on the image of Hankel matrices of (noisy) data have been successfully implemented on real-world systems, e.g., power systems \cite{markovsky2023powersys, huang2019powerconverters}, heating systems \cite{yin2024data, chinde2022data}, quadcopters \cite{elokda2021data} and many others, thus making direct data-based control an important and well-established research topic in modern control theory research.

Essential to \cite{Willems05} and all results that build on it is that the collected data is consecutive or made up of multiple sufficiently long trajectories of the system (see \cite{vanWaarde20}). However, in practice data can be missing due to several reasons, e.g., sensor failure and/or (irregular) packet losses in networked control systems \cite{Sadjadi03}. It might also not be possible to measure some states regularly, e.g., in biomedical applications \cite{bighamian16, Berberich18}. This motivates the development of non-parametric representations of dynamical systems from irregularly measured data, which can then be used for system analysis and control design.

In \cite{Wildhagen23}, the authors use consecutive offline input/state data of discrete-time LTI systems to design stabilizing controllers for closed-loop aperiodically sampled systems. In \cite{Markovsky22b}, assuming availability of consecutive offline data, methods to complete other data sequences with missing samples are proposed by formulating data-driven approximation and interpolation problems. Both of these works still assume the availability of complete data sequences collected offline and, hence, a non-parametric model of the system is assumed to be available (cf. \cite{Willems05}). In contrast, the case where data is (regularly or irregularly) missing in the offline phase is not discussed. The goal of this paper is to exploit the kernel structure of Hankel matrices \cite{Heinig84, Heinig92} in order to retrieve a non-parametric representation of the finite-length behavior of an LTI system using the available (irregularly measured) data.

In the system identification literature, missing data has been addressed in several works. The methods used include frequency domain tools \cite{Pintelon98}, local optimization methods \cite{Markovsky13b} and convex relaxations based on nuclear norm minimization \cite{Liu13}. In this work, we address the problem using the language of the behavioral framework \cite{Willems86}, as was done in \cite{Markovsky13, Markovsky16}. There, a dynamical system is viewed as a set of trajectories, allowing us to define the system properties in terms of its behavior. A representation of the system, however, is defined independently from its behavior and one of several representations of a dynamical system is the kernel representation. An algorithm for identifying a \textit{minimal} kernel representation of scalar systems despite missing data was proposed in \cite{Markovsky13, Markovsky16}. However, no conditions were given to guarantee that this algorithm succeeds for any pattern of missing data.

The contributions of this paper are as follows:
\begin{itemize}
	\item[1.] We first modify the algorithm in \cite{Markovsky13, Markovsky16} and use it to obtain a kernel representation of multivariable systems using irregularly measured data.
	\item[2.] We then exploit the kernel structure of Hankel matrices to recover any \textit{complete} finite-length behavior of the system, i.e., we recover a non-parametric model of the system similar to the ones given by the fundamental lemma \cite{Willems05} and its generalization \cite{Markovsky22}. Different from existing works, this is done using only irregularly measured offline data.
	\item[3.] We consider the special case of periodically missing outputs and provide sufficient conditions on the input such that a kernel representation is guaranteed to be retrieved from the available data. This allows us to propose a fundamental-lemma-like result using data with periodically missing outputs.
	\item[4.] We investigate the presence of noise in the data and provide computational methods to obtain an approximate finite-length behavior of the system.
\end{itemize}

The remainder of the paper is organized as follows: Section~\ref{sec_preliminaries} introduces the notation and reviews preliminary results. Section \ref{sec_problemformulation} formulates the problem of finding kernel representations from missing data, and includes the corresponding algorithm required to compute them. Section \ref{sec_kernelstructure} shows how to use these representations in order to obtain a non-parametric representation of any \textit{complete} finite-length behavior of the system, by exploiting the kernel structure of Hankel matrices. Section \ref{sec_periodicSO} focuses on the special case of periodically missing outputs and provides conditions on the input such that a fundamental-lemma-like result is obtained. In Section~\ref{sec_noisydata} we propose a modification of our results to handle the case of noisy data, such that an approximate finite-length behavior of the system is obtained. Section \ref{sec_examples} verifies the results using numerical examples and illustrates how the resulting representation can be used for approximate data completion in real-world applications, including comparisons to alternative methods. Finally, Section \ref{sec_conclusion} discusses the results and concludes the paper.

%% file: sections/preliminaries.tex
\section{Preliminaries}\label{sec_preliminaries}
The sets of integers, natural and real numbers are denoted by $\mathbb{Z},\mathbb{N},\mathbb{R}$, respectively. The restriction of integers to an interval is denoted by $\mathbb{Z}_{[a,b]}$ for $b>a\in\mathbb{Z}$. We use $I_m$ to denote an $m\times m$ identity matrix and $0_{n\times m}$ to denote an $n\times m$ matrix of zeros; when the dimensions are clear from the context, we omit the subscript for simplicity. We use $\floor{z}$ to denote the floor operator which acts on $z\in\mathbb{R}$ and returns the greatest integer less than or equal to $z$. For a matrix $M\in\mathbb{R}^{m\times n}$, we denote its image by $\mathrm{im}(M)$ and its kernel by $\mathrm{ker}(M)$. When a basis of $\mathrm{ker}(M)$ is to be computed, we write $N=~\mathrm{null}(M)$ which returns a matrix $N$ of appropriate dimensions such that $MN=~0$. The set of polynomial matrices (in $z$) with coefficients in $\mathbb{R}^{m\times n}$ is denoted by $\mathbb{R}^{m\times n}[z]$. For a subspace $\mathcal{A}$ of $\mathbb{R}^n$, we denote its orthogonal complement by $\mathcal{A}^\perp$. Further, for another subspace $\mathcal{B}$ of $\mathbb{R}^n$ such that $\mathcal{A}\cap\mathcal{B}=\{0\}$, the direct sum of the two subspaces is denoted by $\mathcal{A}\oplus\mathcal{B}$.

The set of infinite length $q-$variate, real-valued time series $w=(w(0),w(1),\ldots)$ is denoted by $\left(\mathbb{R}^{q}\right)^\mathbb{N}$. For $T\in\mathbb{N}$, the set of finite-length $q-$variate, real-valued time series $w=(w(0),w(1),\ldots,w(T-1))$ is denoted by $\left(\mathbb{R}^{q}\right)^T$. With slight abuse of notation, we also use $w$ to denote the stacked vector of the time series as
\begin{equation*}
	w=\begin{bmatrix}(w(0))^\top & (w(1))^\top & \cdots & (w(T-1))^\top\end{bmatrix}^\top\in~\mathbb{R}^{qT},
\end{equation*}
and its restriction to $L$ as
\begin{equation*}
	w_{[0,L-1]}=\begin{bmatrix}(w(0))^\top \hspace{-0.5mm}& \hspace{-0.5mm}(w(1))^\top\hspace{-0.5mm} &\hspace{-0.5mm} \cdots\hspace{-0.5mm} & \hspace{-0.5mm}(w(L-1))^\top\end{bmatrix}^\top\hspace{-0.5mm}\in\mathbb{R}^{qL}.
\end{equation*}
The Hankel matrix of depth $L$ of $w$ is defined as
\begin{equation*}
	\mathscr{H}_L(w) = \begin{bmatrix}
			w_{[0,L-1]} & w_{[1,L]} & \cdots & w_{[T-L,T-1]}
		\end{bmatrix}.
	\label{eqn_Hankelmatrix}
\end{equation*}

The behavior $\mathscr{B}$ of a dynamical system is defined as a set of infinite-length trajectories (cf. \cite{Willems86}). The finite-length behavior of a system is a set of finite-length trajectories. For $T\in\mathbb{N}$, this is denoted by $\left.\mathscr{B}\right|_T$ and, hence, a trajectory of length $T$ of the system is denoted by $w\in\left.\mathscr{B}\right|_T\subset\left(\mathbb{R}^q\right)^{T}$. The system is linear if $\mathscr{B}$ is a subspace (of $\left(\mathbb{R}^q\right)^\mathbb{N}$) and it is time-invariant if it is invariant to the action of the \textit{shift operator}, defined as $\sigma:\left(\mathbb{R}^{q}\right)^\mathbb{N}\to\left(\mathbb{R}^{q}\right)^\mathbb{N}$ such that $\sigma^j w(k) \coloneqq w(k+j)$ for $j\in\mathbb{N}$.

Let $u(t)\in\mathbb{R}^m$ and $y(t)\in\mathbb{R}^p$ denote the inputs and outputs of system $\mathscr{B}$ at time $t$. For a permutation matrix $\Pi\in\mathbb{R}^{q\times q}$, we define a partitioning of the variable $w(t)\in\mathbb{R}^q$ such that $w(t)=\Pi\begin{bsmallmatrix} u(t) \\ y(t) \end{bsmallmatrix}$. The set of discrete-time LTI systems with $q-$variables and fixed complexity $(m,n,\ell)$ is denoted by $\partial\mathscr{L}_{m,n,\ell}^{q}$, where $q=m+p$ and $(m,n,\ell)$ denote (i) the number of inputs $\textbf{m}(\mathscr{B})= m$, (ii) the order of the system $\textbf{n}(\mathscr{B})= n$, and (iii) the lag of the system $\boldsymbol{\ell}(\mathscr{B})=\ell$, which is the observability index in the state-space framework. These integers satisfy the following relation $\ell\leq n\leq p\ell$ \cite{Markovsky22}. 

A finite-dimensional LTI system $\mathscr{B}\in\partial\mathscr{L}_{m,n,\ell}^{q}$ admits a kernel representation \cite[Part I]{Willems86}
\begin{equation}
	\mathscr{B} = \mathrm{ ker }~R(\sigma) \coloneqq \{ w:\mathbb{N}\to\mathbb{R}^q ~|~ R(\sigma)w=0\},
	\label{kernel_rep}
\end{equation}
where the operator $R(\sigma)$ is defined by the polynomial matrix
\begin{align}
	R(z) &= R_0 + R_1z + \ldots + R_\ell z^\ell\label{eqn_kernelpolymatrix}\\
	&=\hspace{-1mm} \begin{bmatrix}
		r_1(z)\\
		\vdots\\
		r_g(z)
	\end{bmatrix}\hspace{-1mm} = \hspace{-1mm}\begin{bmatrix}
		r_{1,0} + r_{1,1}z + \ldots + r_{1,\ell_1}z^{\ell_1}\\
		\vdots\\
		r_{g,0} + r_{g,1}z + \ldots + r_{g,\ell_g}z^{\ell_g}
	\end{bmatrix}\hspace{-0.5mm}\in\mathbb{R}^{g\times q}[z],\notag
\end{align}
with $r_{i,j}\in\mathbb{R}^{1\times q}$. A \textit{minimal} kernel representation is one where the degree of the polynomial matrix $R$ is minimal. Each row $r_i(z)$ of $R(z)$ defines an \textit{annihilator} $r(\sigma)$ of the system and is formally defined as follows.
\begin{definition}\textup{\cite[Def. 2]{Markovsky22}}\label{def_annihilator}
	An annihilator of $\mathscr{B}\in\partial\mathscr{L}_{m,n,\ell}^{q}$ is an operator $r(\sigma)$, $r(z)\in\mathbb{R}^{1\times q}[z]$, such that $r(\sigma)\mathscr{B}=0$.
\end{definition}

A finite dimensional LTI system $\mathscr{B}\in\partial\mathscr{L}_{m,n,\ell}^{q}$ also admits an input/state/output representation of the form
\begin{equation}
	\begin{aligned}
		\mathscr{B} &= \mathscr{B}_{ss}(A,B,C,D,\Pi) \coloneqq \left\{\Pi\begin{bsmallmatrix}
			u\\y
		\end{bsmallmatrix}~|~ \exists x\in(\mathbb{R}^n)^\mathbb{N},\right.\\
	&\quad\left.\textup{s.t.}\quad \sigma x = Ax+Bu,\, y=Cx+Du \right\}.
	\end{aligned}\label{eqn_SSrepresentation}
\end{equation}

One of the key ideas in the behavioral framework is the notion of the \textit{most powerful unfalsified model} (MPUM) of the data \cite[Part II]{Willems86}, denoted $\mathscr{B}_{\textup{MPUM}}$. The MPUM always exists and is unique. Its restricted behavior, i.e., $\mathscr{B}_{\textup{MPUM}}|_L$, is given by the image of the Hankel matrix constructed from data. For instance, let $w\in\mathscr{B}|_T$, where $\mathscr{B}\in\partial\mathscr{L}_{m,n,\ell}^{q}$, then $\mathscr{B}_{\textup{MPUM}}|_L=\mathrm{im}(\mathscr{H}_L(w))$. Furthermore, due to linearity and shift-invariance, the following holds $\mathscr{B}_{\textup{MPUM}}|_L=\mathrm{im}(\mathscr{H}_L(w))\subseteq \left.\mathscr{B}\right|_{L}$, where $\left.\mathscr{B}\right|_{L}$ is the restricted behavior of the system. This is summarized in the following theorem, which further gives necessary and sufficient conditions under which the equality holds, i.e., $\mathrm{im}(\mathscr{H}_L(w))= \left.\mathscr{B}\right|_{L}$. Theorem~\ref{thm_identifiability} follows from the generalized persistency of excitation result \cite[Cor. 21]{Markovsky22}, which generalizes the fundamental lemma \cite[Thm. 1]{Willems05} to uncontrollable systems and does not require a priori known input-output partitioning of the data, nor persistency of excitation of the input (see also \cite{Markovsky23pe}).

\begin{theorem}\textup{\cite{Markovsky22}}\label{thm_identifiability}
	Let $w\in\mathscr{B}|_T$ with $\mathscr{B}~\in~\partial\mathscr{L}_{m,n,\ell}^{q}$. Then, for all $L\in\mathbb{Z}_{[\ell,L_{\textup{max}}]}$ where $L_{\textup{max}}\coloneqq\floor*{\frac{T+1}{q+1}}$, the following holds
	\begin{equation}
		\mathrm{im}\left(\mathscr{H}_L(w)\right)\subseteq\mathscr{B}|_L.
	\end{equation}
	Moreover, for $L\geq\ell$, $\mathrm{im}\left(\mathscr{H}_L(w)\right)=\mathscr{B}|_L$ if and only if
	\begin{equation}
		\mathrm{rank}(\mathscr{H}_L(w)) = mL+n.\label{rank_condition_L}\tag{GPE}
	\end{equation}
\end{theorem}

When \eqref{rank_condition_L} holds, one obtains a \textit{non-parametric data-based representation} of the restricted behavior of the system $\mathscr{B}|_L$ for $L\geq\ell$, i.e., any trajectory of the system is given by a linear combination of the columns of the Hankel matrix $\mathscr{H}_L(w)$. This is formalized in the following corollary.
\begin{corollary}\textup{\cite[Cor. 4]{Markovsky22b}}\label{cor_NonParametricRepresentation}
	Let $w\in\mathscr{B}|_T$ with $\mathscr{B}\in\partial\mathscr{L}_{m,n,\ell}^{q}$ and let \eqref{rank_condition_L} hold for $L\geq\ell$. Then $\bar{w}\in\mathscr{B}|_L$ if and only if there exists $\alpha\in\mathbb{R}^{T-L+1}$ such that
	\begin{equation}
		\mathscr{H}_L(w)\alpha = \bar{w}.\label{FL}
	\end{equation}
\end{corollary}

Another result which follows from the rank condition in \eqref{rank_condition_L} is the following. It allows us to retrieve a kernel representation \eqref{eqn_kernelpolymatrix} of an LTI system directly from data. This is formalized in the following corollary.

\begin{corollary}\label{cor_kernelRd}
	Let $w\in\mathscr{B}|_T$ where $\mathscr{B}\in\partial\mathscr{L}_{m,n,\ell}^{q}$. Let $d\geq\ell+1$, and suppose
	\begin{equation}
		\mathrm{rank}(\mathscr{H}_d(w))=md+n.\label{GPE_2}
	\end{equation}
	Then, the coefficients of the polynomial matrix $R$ in \eqref{eqn_kernelpolymatrix} are given by the basis of the left null space of the Hankel matrix $\mathscr{H}_d(w)$, i.e., $R$ is defined in terms of the rows of $R_d\in\mathbb{R}^{pd-n\times qd}$ where
	\begin{equation}
		R_{d} = \begin{bmatrix}
			r_{1,0} & r_{1,1} & \cdots & r_{1,d-1}\\
			r_{2,0} & r_{2,1} & \cdots & r_{2,d-1} \\
			\vdots & \vdots & \ddots & \vdots\\
			r_{pd-n,0} & r_{pd-n,1} & \cdots & r_{pd-n,d-1}
		\end{bmatrix},\label{minimal_annihilator_matrix}
	\end{equation}
	with $r_{i,j}\in\mathbb{R}^{1\times q}$ which satisfies $R_d\mathscr{H}_d(w)=0$.
\end{corollary}
\begin{proof}
	Existence of $R_d$ (of proper dimension) follows from \eqref{GPE_2}. Since \eqref{GPE_2} holds, then im$(\mathscr{H}_d(w))=\mathscr{B}|_d$ (by Theorem \ref{thm_identifiability}). Therefore, the rows of the matrix $R_d$ represent the coefficients of the annihilators of $\mathscr{B}|_d$ (see Definition~\ref{def_annihilator}). Since the map $\mathscr{B}|_d \mapsto \mathscr{B}$ is well defined (see \cite[Lemma 13]{Markovsky22}), then the rows of $R_d$ also represent the coefficients of the annihilators of $\mathscr{B}$ and, hence, define a kernel representation as in \eqref{eqn_kernelpolymatrix} with $g=pd-n$ and $\ell_i=d-1$ for all $i=\{1,\ldots,pd-n\}$.
\end{proof}

One of the goals of this paper is to arrive at a result similar to that of Corollary~\ref{cor_NonParametricRepresentation} when the data is not consecutive (i.e., irregularly measured or contains missing values). To this end, we denote a missing data value by NaN (short for Not a Number). This object is defined such that $0\cdot\textup{NaN}=0$ and $1\cdot\textup{NaN} = \textup{NaN}$. Furthermore, we define an extended set of the real numbers as the union of the set of $\mathbb{R}$ and NaN as $\mathbb{R}_{\textup{ext}}\coloneqq \mathbb{R}\cup\textup{NaN}$. Throughout the paper, we denote a trajectory of $\mathscr{B}|_T$ by $w\in(\mathbb{R}^q)^T$, and its corresponding \textit{irregular measurements}~by $w^{-}\in (\mathbb{R}_{\textup{ext}}^{q})^T$, with NaNs appearing arbitrarily in $w^-$, e.g., $w^{-}(t)=\begingroup\setlength\arraycolsep{1.5pt}\begin{bmatrix} w_1^{-}(t) & \cdots & \textup{NaN} & \cdots & w_q^{-}(t)\end{bmatrix}\endgroup^\top$, for some $t\in\mathbb{Z}_{[0,T-1]}$. This means that we allow for portions of a sample or entire samples of data to be missing. For the majority of the subsequent developments, we consider \textit{exact} data, i.e., noise-free data. Later in Section~\ref{sec_noisydata}, we consider the case of noisy data.

%% file: sections/problemformulation.tex
\section{Kernel representations from missing data}\label{sec_problemformulation}
\subsection{Problem formulation}
Whether in Theorem \ref{thm_identifiability}, Corollary \ref{cor_NonParametricRepresentation} or Corollary \ref{cor_kernelRd}, if the data $w$ is not complete (containing missing values), then one cannot evaluate \eqref{rank_condition_L} or \eqref{GPE_2}, nor use \eqref{FL} as a non-parametric model anymore. One can instead use multiple (short) experiments to build a \textit{mosaic Hankel matrix} (cf. \cite{Markovsky13}). In particular, given $\mathscr{W}=\{w^{(1)},w^{(2)},\ldots,w^{(N)}\}$ with each $w^{(i)}\in\mathscr{B}|_{T_i}$, for $N,T_i\in\mathbb{N}$, $i\in\mathbb{Z}_{[1,N]}$, one can construct
\begin{equation*}
	\mathcal{H}_L(\mathscr{W}) \coloneqq \begin{bmatrix}
		\mathscr{H}_L(w^{(1)}) & \cdots & \mathscr{H}_L(w^{(N)})
	\end{bmatrix},
\end{equation*}
for $\ell\leq L\leq T_i$. Later it was shown in \cite{Markovsky22} that if rank$(\mathcal{H}_L(\mathscr{W}))~=~mL~+~n$, then any $\bar{w}\in\left.\mathscr{B}\right|_L$ if and only if there exists $\alpha~\in~\mathbb{R}^{\sum_{i=1}^{N} (T_i - L+1)}$ such that $\mathcal{H}_L(\mathscr{W})\alpha = \bar{w}$. Similar ideas were presented in \cite{vanWaarde20}, but only controllable systems were addressed, an a priori known input-output partitioning was required and a collective persistency of excitation condition on the input sequences $u^{(i)}$ was imposed. However, both \cite{Markovsky22, vanWaarde20} require that each trajectory $w^{(i)}$ is composed of at least $L$ consecutive samples.

In this work, we provide computational methods to retrieve a non-parametric model as in \eqref{FL} from data which has (regularly or irregularly) missing values. We do not require that data is consecutive for $L$ samples, nor do we impose controllability or persistency of excitation. To do so, we modify the algorithm in \cite{Markovsky13, Markovsky16} to compute a (non-minimal) kernel representation of the system. For convenience, we first illustrate the procedure with the following (simple) motivating example and later formalize the algorithm in Section~\ref{sec_algorithm}. Note that none of the following steps have to be carried out by the user, instead all steps are implemented in a single algorithm (see Algorithm~\ref{alg_generalalgorithm}).
\begin{example}\label{example_MotivatingExample}
	Consider the system $w(t)=2w(t-1)-w(t-2)$. Here, $n=2,\,\ell=2,\,p=1,\,m=0$. Suppose an experiment of length $T=8$ was performed on the system from initial conditions $w(0)=1$ and $w(1)=2$. The following are the complete and missing data sequences, respectively
	\begin{equation*}
		\begin{aligned}
			w &= \{1,2,3,4,5,6,7,8\},\\
			w^{-} &= \{1,2,\textup{NaN},4,5,\textup{NaN},7,8\}.
		\end{aligned}
	\end{equation*}
	The data is missing periodically with period equal to $\ell+1$. This prevents us from obtaining the kernel representation as in Corollary \ref{cor_kernelRd} since
	\begin{equation*}
		\mathscr{H}_{\ell+1}(w^{-}) = \begingroup\setlength\arraycolsep{2pt}\begin{bmatrix}
			1 & 2 & \textup{NaN} & 4 & 5 & \textup{NaN}\\
			2 & \textup{NaN} & 4 & 5 & \textup{NaN} & 7\\
			\textup{NaN} & 4 & 5 & \textup{NaN} & 7 & 8
		\end{bmatrix}\endgroup
	\end{equation*}
	has missing values in every row and every column. Instead, one constructs an extended series by concatenating $w$ with $T-$long instances of \textup{NaN}, i.e.,
	\begin{equation*}
		w_{\textup{ext}} = (\underbrace{1,2,\textup{NaN},4,5,\textup{NaN},7,8}_{w^{-}},\underbrace{\textup{NaN},\cdots,\textup{NaN}}_{T\textup{ times}}),
	\end{equation*}
	where $T=8$. This extended vector allows us to investigate the following Hankel matrix, which encompasses all other shallower Hankel matrices of $w^-$
	\begin{align}
			&\mathscr{H}_T(w_{\textup{ext}}) = \label{eqn_FullHankelMat}\\
			&\begin{bmatrix}
				\begingroup\setlength\arraycolsep{3.5pt}\begin{array}{c|c}
					\begin{matrix}
						1 & 2 & \textup{NaN} & 4 & 5\\
						2 & \textup{NaN} & 4 & 5 & \textup{NaN}\\
						\textup{NaN} & 4 & 5 & \textup{NaN} & 7\\
						4 & 5 & \textup{NaN} & 7 & 8
					\end{matrix} & \begin{matrix}
						\textup{NaN} & 7 & 8\\
						7 & 8 & \textup{NaN}\\
						8 & \textup{NaN} & \textup{NaN}\\
						\textup{NaN} & \textup{NaN} & \textup{NaN}\\
					\end{matrix} \\ \hline 
					\begin{matrix}
						5 & \textup{NaN} & 7 & 8 & \textup{NaN}\\
						\textup{NaN} & 7 & 8 & \textup{NaN} & \textup{NaN}\\
						7 & 8 & \textup{NaN} & \textup{NaN} & \textup{NaN}\\
						8 & \textup{NaN} & \textup{NaN} & \textup{NaN} & \textup{NaN}
					\end{matrix} & \begin{matrix}
						\textup{NaN} & \textup{NaN} & \textup{NaN}\\
						\textup{NaN} & \textup{NaN} & \textup{NaN}\\
						\textup{NaN} & \textup{NaN} & \textup{NaN}\\
						\textup{NaN} & \textup{NaN} & \textup{NaN}
					\end{matrix}
				\end{array}\endgroup
			\end{bmatrix}.\notag
	\end{align}

	As in Corollary \ref{cor_kernelRd}, when $\mathrm{rank}(\mathscr{H}_{d}(w))=md+n$, the kernel representation of the system is given by the rows of $R_d$ where $R_{d}\mathscr{H}_{d}(w)=0$ for some $d\geq\ell+1$. The key step is to select submatrices of $\mathscr{H}_{d}(w^{-})$ that do not have \textup{NaN} values but have non-trivial left kernels. For instance, consider
	\begin{equation*}
		{H}_1 = \begin{bmatrix}
			1 & 4\\
			2 & 5\\
			\textup{NaN} & \textup{NaN}\\
			4 & 7
		\end{bmatrix},\, H_2 = \begin{bmatrix}
			2 & 5\\
			\textup{NaN} & \textup{NaN}\\
			4 & 7\\
			5 & 8
		\end{bmatrix}, \, H_3=\begin{bmatrix}
		\textup{NaN}\\4\\5\\\textup{NaN}
	\end{bmatrix}\hspace{-1mm},
	\end{equation*}
	which are found by taking the first and fourth, second and fifth, and third columns of $\mathscr{H}_{\ell+2}(w^{-})$, respectively (here $d=\ell+2$ and $\mathscr{H}_{\ell+2}(w^{-})$ is the indicated top left submatrix of \eqref{eqn_FullHankelMat}). By deleting the shared \textup{NaN} rows of $H_1,H_2,H_3$, one can find bases for the non-trivial left kernels of the resulting matrices $\bar{H}_1,\bar{H}_2,\bar{H}_3$ as
	\begin{equation*}
		\begin{aligned}
			&\underbrace{\begin{bmatrix}
					1 & -3/2 & 1/2
			\end{bmatrix}}_{\bar{Z}_1}\underbrace{\begin{bmatrix}
					1 & 4\\
					2 & 5\\
					4 & 7
			\end{bmatrix}}_{\bar{H}_1}=0,\quad \underbrace{\begin{bmatrix}
					1 & -3 & 2
			\end{bmatrix}}_{\bar{Z}_2}\underbrace{\begin{bmatrix}
					2 & 5\\
					4 & 7\\
					5 & 8
			\end{bmatrix}}_{\bar{H}_2}=0,\\
			&\underbrace{\begin{bmatrix}
					-5/4 & 1
			\end{bmatrix}}_{\bar{Z}_3}\underbrace{\begin{bmatrix}
			4\\5
		\end{bmatrix}}_{\bar{H}_3}=0.
		\end{aligned}
	\end{equation*}
	When extending $\bar{Z}_i$ to $Z_i$ by inserting zeros in place of the deleted \textup{NaN} rows, the following holds by definition of \textup{NaN}
	\begin{equation*}
		\begin{aligned}
			&\underbrace{\begingroup\setlength\arraycolsep{2.5pt}\begin{bmatrix}
					1 & -3/2 & 0 & 1/2
				\end{bmatrix}\endgroup}_{Z_1}\begingroup\setlength\arraycolsep{2pt}\begin{bmatrix}
				1 & 4\\
				2 & 5\\
				\textup{NaN} & \textup{NaN}\\
				4 & 7
			\end{bmatrix}\endgroup=0,\\
			&\underbrace{\begingroup\setlength\arraycolsep{2.5pt}\begin{bmatrix}
					1 & 0 & -3 & 2
				\end{bmatrix}\endgroup}_{Z_2}\begingroup\setlength\arraycolsep{2pt}\begin{bmatrix}
				2 & 5\\
				\textup{NaN} & \textup{NaN}\\
				4 & 7\\
				5 & 8
			\end{bmatrix}\endgroup=0,\quad\underbrace{\begingroup\setlength\arraycolsep{2.5pt}\begin{bmatrix}
					0 &-5/4 &1 &0
				\end{bmatrix}\endgroup}_{Z_3}\begin{bmatrix}
				\textup{NaN}\\ 4\\ 5\\ \textup{NaN}
			\end{bmatrix}\hspace{-1mm}=\hspace{-1mm}0.
		\end{aligned}
	\end{equation*}
	Consider a vector $\bar{Z}_i$ in the left kernel of a submatrix $\bar{H}_i$. In order for the extended vector $Z_i$ to be in the left kernel of $\mathscr{H}_{\ell+2}(w)$, we require that $\bar{H}_i$ and $\mathscr{H}_{\ell+2}(w)$ have the same rank (see Lemma \ref{lemma_submatrices} below). In particular,
	\begin{gather}
		\hspace{-38mm}\mathrm{rank}(\bar{H}_i)=\mathrm{rank}(\mathscr{H}_{\ell+2}(w))  \implies \label{eqn_RankConditionOnSubmatrices}\\
		\hspace{20mm}\left(\bar{Z}_i^\top\in\mathrm{ker}(\bar{H}_i^\top) \implies Z_i^\top\in\mathrm{ker}(\mathscr{H}_{\ell+2}^\top(w))\right).\notag
	\end{gather}

	We point out that $\mathrm{rank}(\mathscr{H}_{\ell+2}(w))$ cannot be evaluated since $w$ is not completely known. However, for existence of $R_{\ell+2}$ we require that  $\mathrm{rank}(\mathscr{H}_{\ell+2}(w))=m(\ell+2)+n$ (compare Corollary~\ref{cor_kernelRd}). Since the complexity is known, one can test whether $\mathrm{rank}(\bar{H}_i)=m(\ell+2)+n$ as the matrices $\bar{H}_i$ do not have missing values.
	
	For this example, $\mathrm{rank}(\bar{H}_1)=\mathrm{rank}(\bar{H}_2)=m(\ell+2)+n=2$, whereas $\mathrm{rank}(\bar{H}_3)=1$. From \eqref{eqn_RankConditionOnSubmatrices}, it holds that $\mathscr{Z} \mathscr{H}_{\ell+2}(w)=0$, where
	\begin{equation}
		\mathscr{Z}=\begin{bmatrix}
			Z_1 \\ Z_2
		\end{bmatrix} = \begin{bmatrix}
		1 & -3/2 & 0 & 1/2\\
		1 & 0 & -3 & 2
	\end{bmatrix}.
	\end{equation}
	
	Finally, a basis for the kernel representation $R_{\ell+2}$ is given by the $p(\ell+2)-n=2$ linearly independent rows in $\mathscr{Z}$. For this example, $\mathscr{Z}=R_{\ell+2}$.
\end{example}
The steps followed in Example \ref{example_MotivatingExample} are a modification of the results of \cite{Markovsky13}. Unlike \cite{Markovsky13}, our method finds a (not necessarily minimal) kernel representation and works for multivariable systems. The most significant modification was to impose additional rank constraints on the selected submatrices (see \eqref{eqn_RankConditionOnSubmatrices} and Lemma~\ref{lemma_submatrices} below) such that the output of the algorithm is indeed a kernel representation of the system. This is an important condition for the selected submatrices to be \textit{informative} in the sense that the left kernel of each submatrix reveals a part of the kernel representation of $\mathscr{B}$ (compare \eqref{rank_condition_L} and Corollary~\ref{cor_kernelRd}). In the next subsection, we formalize the steps of Example~\ref{example_MotivatingExample}. Later in Section~\ref{sec_periodicSO}, we consider a special case of periodically missing outputs and provide conditions on the input such that the algorithm is guaranteed, a priori, to return the kernel representation.
\subsection{Algorithm for computing the kernel representation}\label{sec_algorithm}
In this subsection, we formalize the steps followed in Example \ref{example_MotivatingExample}. We do not make any assumption on the pattern of missing data, and only assume knowledge of the model complexity. We start by stating the following lemma, which formalizes the rank condition \eqref{eqn_RankConditionOnSubmatrices}. This lemma is needed in our modification of the algorithm in \cite{Markovsky13} (see Algorithm \ref{alg_generalalgorithm} below). Specifically, it represents a condition on the selected submatrix to be informative in the sense that its left kernel reveals a part of the system's annihilators.

\begin{lemma}\label{lemma_submatrices}
	Let $A\in\mathbb{R}^{n\times m}$ be a block matrix of the form
	\begin{equation*}
		A = \begin{bmatrix}
			A_{11} & A_{12}\\ A_{21} & A_{22}
		\end{bmatrix},
	\end{equation*}
	with $A_{11}\in\mathbb{R}^{n_1\times m_1},\,A_{12}\in\mathbb{R}^{n_1\times m_2},\,A_{21}\in\mathbb{R}^{n_2\times m_1}$, $A_{22}\in\mathbb{R}^{n_2\times m_2}$ and where $n=n_1+n_2$ and $m=m_1+m_2$. Suppose $A$ has a non-trivial left kernel, i.e., $\mathrm{rank}(A)<n$. If $\mathrm{rank}(A_{11}) = \mathrm{rank}(A)$ and $A_{11}$ has a non-trivial left kernel, then
	\begin{equation}
		\bar{z} A_{11}=0\implies z A=0, \quad \forall \bar{z}^\top\in\mathrm{ker}(A_{11}^\top)
	\end{equation}
	where $z=\begin{bmatrix}
		\bar{z} & 0_{1\times n_2}
	\end{bmatrix}\in\mathbb{R}^{1\times n}$.
\end{lemma}
\begin{proof}
	Since $\mathrm{rank}(A_{11}) = \mathrm{rank}(A)$ and $A_{11}$ has a non-trivial left kernel, there exists $\bar{z}^\top\in\mathbb{R}^{n_1}$ such that $\bar{z} A_{11}=0$. As in the lemma statement, let $z=\begin{bmatrix}
		\bar{z} & 0_{1\times n_2}
	\end{bmatrix}$. By construction, it holds that
	\begin{equation}
		z \begin{bmatrix}
			A_{11} \\ A_{21}
		\end{bmatrix} = \begin{bmatrix}
			\bar{z} & 0_{1\times n_2}
		\end{bmatrix}\begin{bmatrix}
			A_{11} \\ A_{21}
		\end{bmatrix} = \bar{z} A_{11} + 0\cdot A_{21} = 0.\label{eqn_ProofKernelLemma1}
	\end{equation}%
	Furthermore, since $\mathrm{rank}(A_{11}) = \mathrm{rank}(A)$, it holds that the columns of $A_{12}$ can be written as a linear combination of the columns of $A_{11}$. This means that there exists a matrix $K\in\mathbb{R}^{m_1\times m_2}$ such that $A_{12}=A_{11}K$ and, hence,
	\begin{equation}
		z \begin{bmatrix}
			A_{12} \\ A_{22}
		\end{bmatrix} = \begin{bmatrix}
			\bar{z} & 0_{1\times n_2}
		\end{bmatrix}\begin{bmatrix}
			A_{11}K \\ A_{22}
		\end{bmatrix} = \bar{z} A_{11}K + 0\cdot A_{22} = 0.\label{eqn_ProofKernelLemma2}
	\end{equation}
	
	Equations \eqref{eqn_ProofKernelLemma1} and \eqref{eqn_ProofKernelLemma2} together imply that $z A = 0$, which completes the proof.
\end{proof}
\begin{remark}\label{remark_KernelLemma}
	The rows and columns of $A_{ij}$ in Lemma \ref{lemma_submatrices} need not appear consecutively in the matrix $A$. This is because row and column permutations do not change the rank of $A$. To see how Lemma \ref{lemma_submatrices} generalizes \eqref{eqn_RankConditionOnSubmatrices} in Example \ref{example_MotivatingExample}, one can carry out the necessary row/column permutations that bring the submatrix $\bar{H}_i$ in $\mathscr{H}_{\ell+2}(w^-)$ to the top left corner, i.e., the location of $A_{11}$ in the matrix $A$ in Lemma \ref{lemma_submatrices}.
\end{remark}
Algorithm \ref{alg_generalalgorithm} generalizes the steps followed in Example~\ref{example_MotivatingExample}. The algorithm takes as an input a sequence of irregularly measured data $w^{-}$ and does not require any further steps to be performed by the user. Then, it searches over different depths of $\mathscr{H}_T(w_{\textup{ext}})$ for possible submatrices $\bar{H}_i$ that have non-trivial left kernels (checkable by a rank condition). According to Lemma~\ref{lemma_submatrices} (compare also \eqref{eqn_RankConditionOnSubmatrices} and Remark~\ref{remark_KernelLemma}), suitable extensions of vectors in the left null space of the aforementioned $\bar{H}_i$ are guaranteed to form annihilators of the system. These annihilators are collected in the matrix\footnote{Notice that the matrix $\mathscr{Z}$ is adapted at each iteration such that information from previous iterations is taken into account.} $\mathscr{Z}$. If enough annihilators are found, or if the maximum depth of the Hankel matrix is reached, the algorithm terminates. The algorithm is deemed successful if $\mathscr{Z}$ contains $pd-n$ linearly independent rows (where $pd-n$ is the dimension of the left kernel of $\mathscr{H}_d(w)$). These rows specify a kernel representation of the system (cf. Corollary~\ref{cor_kernelRd}). If $\mathrm{rank}(\mathscr{Z})<pd-n$, then the algorithm did not succeed in returning the kernel representation.%
\begin{algorithm}[!t]
	\caption{Computing a kernel representation}\label{alg_generalalgorithm}
	\textbf{Input:} Irregular measurements $w^-\in(\mathbb{R}_{\textup{ext}}^{q})^T$ of a trajectory $w\in\mathscr{B}|_T$ where $\mathscr{B}\in\partial\mathscr{L}_{m,n,\ell}^{q}$ is of known complexity.
	\begin{itemize}
		\item[1.] Construct the extended sequence $w_{\textup{ext}}$ by appending $w^{-}$ with $T$ instances of NaN.
		\item[2.] Construct the matrix $\mathscr{H}_T(w_{\textup{ext}})$, and initialize $\mathscr{Z}=[\,]$.
		\item[3.] \textbf{For} $\ell+1\leq d\leq T$ \textbf{do}
		\begin{itemize}
			\item Search over all submatrices of $\mathscr{H}_d(w^{-})$, i.e., $\bar{H}_i\in\mathbb{R}^{\eta_i\times \rho_i}$, that have $\mathrm{rank}(\bar{H}_i)=md+n$ and non-trivial left kernels i.e., $\eta_i>md+n$.
			\item Compute a basis for $\mathrm{ker}(\bar{H}_i^\top)$, i.e., a full row rank matrix $\bar{Z}_i\in\mathbb{R}^{\zeta_i\times\eta_i}$ such that $\bar{Z}_i \bar{H}_i=0$.
			\item Extend $\bar{Z}_i$ to $Z_i$ by inserting zero columns at the location of the rows removed from $\mathscr{H}_d(w^-)$ to obtain $\bar{H}_i$.
			\item Concatenate all the above matrices $Z_i$ in $\mathscr{Z}$, i.e., $\mathscr{Z}\coloneqq\begin{bmatrix}
					[\begin{matrix}
						\mathscr{Z} & \mathbf{0}
					\end{matrix}]^\top & Z_1^\top & Z_2^\top & \cdots
				\end{bmatrix}^\top$, where $\mathbf{0}$ denotes a matrix of zeros of appropriate dimensions.
			\item \textbf{If} $\mathrm{rank}(\mathscr{Z})=pd-n$ \textbf{break}.
		\end{itemize}
		\textbf{end for}
	\end{itemize}
	\textbf{Output:} The matrix $R_d$ specifying the kernel representation is given by any $pd-n$ linearly independent rows of $\mathscr{Z}$.
\end{algorithm}%

At this point, we do not impose any conditions on the pattern of the missing data. As a result, we cannot a priori guarantee whether or not this algorithm will succeed or fail. The algorithm may fail, for instance, if there are not enough submatrices $\bar{H}_{i}$ which have the desired rank (i.e., data is not informative). However, even if the algorithm fails, it still reveals a part of the true system representation given by the annihilators which the algorithm manages to find. In general, the success rate of Algorithm~\ref{alg_generalalgorithm} increases as the length of the sequence $T$ increases and as the fraction of given to missing samples increases as well. This is illustrated by an example in Section~\ref{sec_ExA}.

In some cases, prior knowledge on the pattern of missing data is available and one can exploit it to guarantee successful completion of Algorithm~\ref{alg_generalalgorithm}. In Section~\ref{sec_periodicSO} we consider the case of periodically missing output samples and show sufficient conditions (including a PE condition on the input and a lower bound for the length of the sequence $T$) such that Algorithm~\ref{alg_generalalgorithm} is guaranteed to return a kernel representation of the system.

We point out that the most expensive step of Algorithm~\ref{alg_generalalgorithm} is the rank check and computation of a basis for the left kernel of submatrices $\bar{H}_i$. This corresponds to a singular value decomposition of that matrix, which has complexity $O\big(\max\{\eta_i,\rho_i\}(\min\{\eta_i,\rho_i\})^2)$ (cf. \cite{trefethen2022}), where $\eta_i$ and $\rho_i$ are the dimensions of the complete submatrix $\bar H_i$ of the Hankel matrix $\mathscr{H}_d(w^-)$. Later in Section~\ref{sec_examples}, we illustrate with numerical examples that Algorithm~\ref{alg_generalalgorithm} takes less time to complete compared to other methods that rely on data completion (which requires solving nonlinear or convex programs, e.g., convex relaxations based on nuclear norm minimization \cite{Liu13}) followed by finding a basis for the left kernel.

In the following section, we use the output of Algorithm~\ref{alg_generalalgorithm} to obtain the \textit{complete} finite-length behavior of the system $\mathscr{B}|_L$, for \textit{any} $L\geq d$.

%% file: sections/kernelstructure.tex
\section{Retrieving a non-parametric system representation}\label{sec_kernelstructure}
Once a kernel representation of an LTI system is obtained, e.g., by Algorithm \ref{alg_generalalgorithm}, we use this kernel representation to reconstruct the restricted behavior of the system. This can then be used as a non-parametric model of the system similar to that in Corollary~\ref{cor_NonParametricRepresentation}. We do this by exploiting the kernel structure of Hankel matrices \cite{Heinig84, Heinig92}.
\subsection{Kernel structure of Hankel matrices}
We start by reviewing the concept of \textit{shift-chains}.
	\begin{definition}[\cite{Heinig84}]\label{def_shiftchains}
		A shift-chain of length $r$ generated by a vector $z\in\mathbb{R}^s$ with $z\neq0$ is a set of vectors $v_0,\ldots,v_{r-1}\in\mathbb{R}^n$, with $s+r=n+1$, such that $v_k = \Pi_{n,s}^{k} z,\,\forall k\in\mathbb{Z}_{[0,r-1]}$ where $\Pi_{n,s}^k$ is a permutation matrix having $n$ rows, $s$ columns and $k$ specifying the non-zero diagonal.
\end{definition}

To better illustrate the definition of a shift-chain, consider the following example.
	\begin{example}
		Let $z=\begin{bmatrix}
			a & b & c
		\end{bmatrix}^\top\in\mathbb{R}^3$ and let $n=5$ (hence, $r=n+1-s=3$). The following vectors form a shift-chain
		\begin{equation*}
			\begin{aligned}
				v_0 &= \begin{bmatrix}
					a& b& c& 0& 0
				\end{bmatrix}^\top,\quad v_1 = \begin{bmatrix}
				0& a& b& c& 0
			\end{bmatrix}^\top,\\
				v_2 &= \begin{bmatrix}
					0& 0& a& b& c
				\end{bmatrix}^\top.
			\end{aligned}
		\end{equation*}
\end{example}

In the work by Heinig et al. \cite{Heinig84, Heinig92}, the kernel structure of Hankel matrices were investigated. Being a structured matrix, its kernels posses certain structures as well. The following theorem shows how the kernels of a Hankel matrix can be represented using shift-chains.

\begin{theorem}\textup{\cite[Cor. 2.3]{Heinig92}}\label{thm_Kernelstructure2}
	For a sequence $t=(t_0,t_1,\ldots)$, where $t_i\in\mathbb{R}^{p\times q}$, the (left or right) kernel of a Hankel matrix of $t$ is given by the linear hull of at most $p+q$ shift-chains.
\end{theorem}

In the next subsection, we construct a shift-chain-like matrix out of the computed kernel representation obtained from Algorithm \ref{alg_generalalgorithm} and use it to later obtain the \textit{complete} restricted behavior (of \textit{any} length $L\geq d\geq\ell+1$) of the LTI system. This is done by showing that the shift-chain-like matrix forms a basis for the left kernel of any deeper Hankel matrix (whose image is equal to $\mathscr{B}|_L$) and then using orthogonality of the associated subspaces of the Hankel matrix to retrieve the restricted behavior of the system.

\subsection{Retrieving the restricted behavior of an LTI system}
In Section \ref{sec_algorithm} we used Algorithm \ref{alg_generalalgorithm} to obtain a kernel representation of the system using irregularly measured data. If successful, the algorithm returns a full row rank matrix $R_d\in\mathbb{R}^{pd-n\times qd}$ such that $R_d\mathscr{H}_d(w)=0$. In contrast, Theorem \ref{thm_identifiability} provides a non-parametric model of the restricted behavior of an LTI system as the image of a Hankel matrix consisting of the data generated by the system $\mathscr{H}_L(w)$, provided that this matrix satisfies the rank condition \eqref{rank_condition_L} (compare \cite{Willems05, Markovsky22}). In this subsection, we use a kernel representation $R_d$ to find a complete matrix (i.e., without missing values) that has the same rank and image of the matrix $\mathscr{H}_L(w)$, thus corresponding to a non-parametric model of the restricted behavior of the true data-generating system. Furthermore, we show how one can retrieve the \textit{complete} restricted behavior of the system $\mathscr{B}|_L$ for \textit{any} $L\geq d\geq\ell+1$, i.e., \textit{independent of the number of available data points}. We start with an illustrative example using the same system as in Example~\ref{example_MotivatingExample}. 
\begin{example}\label{example_Bmatrix}
	Consider again the system and setting of Example~\ref{example_MotivatingExample}. There, it was shown that one can use $w^-$ to obtain a basis for the left kernel of $\mathscr{H}_{\ell+2}(w)$, i.e., a matrix $R_{\ell+2}$ which satisfies $R_{\ell+2}\mathscr{H}_{\ell+2}(w)={0}$ and takes the form
	\begin{equation*}
		\begin{aligned}
			R_{\ell+2}=\begingroup\setlength\arraycolsep{2pt}\begin{bmatrix}
				r_{1,0} & r_{1,1} & r_{1,2} & r_{1,3}\\
				r_{2,0} & r_{2,1} & r_{2,2} & r_{2,3}
			\end{bmatrix}\endgroup=\begingroup\setlength\arraycolsep{2pt}\begin{bmatrix}
				1 & -3/2 & 0 & 1/2\\
				1 & 0 & -3 & 2
			\end{bmatrix}\endgroup.
		\end{aligned}
	\end{equation*}
	By Corollary~\ref{cor_kernelRd}, $R_{\ell+2}$ specifies a kernel representation of the system. This means that for any other trajectory of the same system $\bar{w}\in\mathscr{B}_{\bar{T}}$, with $\bar{T}\geq 8$, $R_{\ell+2}\mathscr{H}_{\ell+2}(\bar{w})={0}$ holds, i.e.,
	\begin{equation}
		\begingroup\setlength\arraycolsep{1.5pt}\begin{bmatrix}
			r_{1,0} & r_{1,1} & r_{1,2} & r_{1,3}\\
			r_{2,0} & r_{2,1} & r_{2,2} & r_{2,3}
		\end{bmatrix}\endgroup\begingroup\setlength\arraycolsep{1.75pt}\begin{bmatrix}
			\bar{w}(0) & \bar{w}(1) & \cdots & \bar{w}({\bar{T}-4})\\
			\bar{w}(1) & \bar{w}(2) & \cdots & \bar{w}({\bar{T}-3})\\
			\bar{w}(2) & \bar{w}(3) & \cdots & \bar{w}({\bar{T}-2})\\
			\bar{w}(3) & \bar{w}(4) & \cdots & \bar{w}({\bar{T}-1})
		\end{bmatrix}\endgroup={0}.\label{eqn_anynewtrajectory}
	\end{equation}
	Let $\bar{T}=9$, then one can construct the following matrix
	\begin{equation}
		\mathscr{H}_{\ell+3}(\bar{w})=\begin{bmatrix}
			\bar{w}(0) & \bar{w}(1) & \cdots & \bar{w}(4)\\
			\bar{w}(1) & \bar{w}(2) & \cdots & \bar{w}(5)\\
			\bar{w}(2) & \bar{w}(3) & \cdots & \bar{w}(6)\\
			\bar{w}(3) & \bar{w}(4) & \cdots & \bar{w}(7)\\
			\bar{w}(4) & \bar{w}(5) & \cdots & \bar{w}(8)
		\end{bmatrix}.
	\end{equation}
	Define the following shift-chain-like matrix
	\begin{equation}
		\begin{aligned}
		\Gamma &= \begin{bmatrix}
			r_{1,0} & r_{1,1} & r_{1,2} & r_{1,3} & 0\\
			r_{2,0} & r_{2,1} & r_{2,2} & r_{2,3} & 0\\
			0 & r_{1,0} & r_{1,1} & r_{1,2} & r_{1,3}
		\end{bmatrix},\label{eqn_Bmatrix}
		\end{aligned}
	\end{equation}
	which is constructed from $R_{\ell+2}$ by appending each row by $m+p$ zeros (first two rows of $\Gamma$), then taking the first $p$ rows of $R_{\ell+2}$ and shifting them by $m+p$ zeros (last row of $\Gamma$). Notice that, by \eqref{eqn_anynewtrajectory}, it holds that $\Gamma\mathscr{H}_{\ell+3}(\bar{w})={0}$. Furthermore, if $\mathrm{rank}(\mathscr{H}_{\ell+3}(\bar{w}))=m(\ell+3)+n$, then the rows of $\Gamma$ represent a basis for the left kernel of $\mathscr{H}_{\ell+3}(\bar{w})$.
\end{example}

In the following, we formalize the procedure followed in Example~\ref{example_Bmatrix}. Recall that, if successful, Algorithm~\ref{alg_generalalgorithm} returns a full row rank matrix $R_d\in\mathbb{R}^{pd-n\times qd}$ such that $R_d \mathscr{H}_d(w)={0}$. Given such a matrix, we define, for any $L\geq d\geq \ell+1$, the following shift-chain-like matrix $\Gamma\in\mathbb{R}^{pL-n\times qL}$
\begin{align}
	&\Gamma =\label{Bmatrix}\\
	&\notag\begin{tikzpicture}[decoration={brace,amplitude=5pt},baseline=(current bounding box.west), scale=0.95, every node/.style={scale=0.9}]
		\node[align=center] at (0,0) {$\begingroup\setlength\arraycolsep{3pt}\begin{bmatrix}
				\begin{matrix}
					r_{1,0}\\
					r_{2,0}\\
					\vdots\\
					r_{pd-n,0}
				\end{matrix} & \begin{matrix}
					r_{1,1}\\
					r_{2,1}\\
					\vdots\\
					r_{pd-n,1}
				\end{matrix} & \begin{matrix}
					\cdots\\ \cdots\\ \ddots\\ \cdots
				\end{matrix} & \begin{matrix}
					r_{1,d-1}\\
					r_{2,d-1}\\
					\vdots\\
					r_{pd-n,d-1}
				\end{matrix} &\\
				& \begin{matrix}
					r_{1,0}\\
					r_{2,0}\\
					\vdots\\
					r_{p,0}
				\end{matrix} & \begin{matrix}
					r_{1,1}\\
					r_{2,1}\\
					\vdots\\
					r_{p,1}
				\end{matrix} & \begin{matrix}
					\cdots\\ \cdots\\ \ddots\\ \cdots
				\end{matrix} & \begin{matrix}
					r_{1,d-1}\\
					r_{2,d-1}\\
					\vdots\\
					r_{p,d-1}
				\end{matrix}\\
				& &\ddots & \ddots & \ddots & \ddots\\
				& & & \begin{matrix}
					r_{1,0}\\
					r_{2,0}\\
					\vdots\\
					r_{p,0}
				\end{matrix} & \begin{matrix}
					r_{1,1}\\
					r_{2,1}\\
					\vdots\\
					r_{p,1}
				\end{matrix} & \begin{matrix}
					\cdots\\ \cdots\\ \ddots\\ \cdots
				\end{matrix} & \begin{matrix}
					r_{1,d-1}\\
					r_{2,d-1}\\
					\vdots\\
					r_{p,d-1}
				\end{matrix}
			\end{bmatrix}\endgroup$};
		\draw[decorate] (2,1.25) -- (3.5,-1) node[above=5pt,midway,sloped] {$L-d$ times};
	\end{tikzpicture}.
\end{align}
By construction, the matrix $\Gamma$ has full row rank, i.e., $\mathrm{rank}(\Gamma)=pL-n$. This is because (i) the rows of $R_d$ are linearly independent and (ii) the structure of $\Gamma$ where each row is appended by an appropriate number of zeros.

For any trajectory $\bar{w}\in\mathscr{B}|_{\bar{T}}$ for which $\mathrm{rank}(\mathscr{H}_L(\bar{w}))=mL+n$ (and hence the corresponding $\mathscr{H}_L(\bar{w})$ represents a non-parametric representation of the system as in Corollary~\ref{cor_NonParametricRepresentation}), the following lemma states that the matrix $\Gamma$ in \eqref{Bmatrix} constitutes a basis for the left kernel of the Hankel matrix $\mathscr{H}_L(\bar{w})$, i.e., $\Gamma\mathscr{H}_L(\bar{w})={0}$. Note that $\bar{w}$ need not be measured for the results to hold. However, existence of such a trajectory is guaranteed by the fact that dim$(\mathscr{B}|_L)=mL+n$ (see Theorem~\ref{thm_identifiability} and \cite{Markovsky22} for more details).
\begin{lemma}
	Consider an LTI system $\mathscr{B}\in\partial\mathscr{L}_{m,n,\ell}^{q}$ and let $w\in\mathscr{B}|_T$. Given a full row rank matrix $R_d$ satisfying $R_d\mathscr{H}_d(w)=~0$, then for any $L\geq d\geq\ell+1$ and any other $\bar{w}\in\mathscr{B}|_{\bar{T}}$ satisfying $\mathrm{rank}(\mathscr{H}_L(\bar{w}))=mL+n$, it holds that the rows of the matrix $\Gamma$ in \eqref{Bmatrix} represent a basis for the left kernel of $\mathscr{H}_L(\bar{w})$, i.e., $\Gamma \mathscr{H}_L(\bar{w})=0$ and $\mathrm{rank}(\Gamma)=\mathrm{dim}(\mathrm{ker}(\mathscr{H}^\top_L(\bar{w})))=pL-n$.
\end{lemma}
\begin{proof}
	According to Corollary~\ref{cor_kernelRd}, the matrix $R_d$ specifies a kernel representation of the system. This means that for $\bar{w}\in\mathscr{B}|_{\bar{T}}$, the following holds
	\begin{equation}
		R_d\mathscr{H}_d(\bar{w})=0.\label{eqn_selectedwindows}
	\end{equation}
	It is assumed that $\mathrm{rank}(\mathscr{H}_L(\bar{w}))=mL+n$ which implies that the dimension of its left kernel is $\mathrm{dim}(\mathrm{ker}(\mathscr{H}^\top_L(\bar{w})))=pL-n=\mathrm{rank}(\Gamma)$, where the rank of the matrix $\Gamma$ follows from the discussion below \eqref{Bmatrix}. The proof is concluded by noticing that the product of the non-zero elements of $\Gamma$ with $\mathscr{H}_L(\bar{w})$ involves selected windows of \eqref{eqn_selectedwindows} which, along with the remaining zero elements of $\Gamma$, results in the zero matrix, i.e., $\Gamma\mathscr{H}_L(\bar{w})={0}$. Since $\mathrm{rank}(\Gamma)=\mathrm{dim}(\mathrm{ker}(\mathscr{H}^\top_L(\bar{w})))=pL-n$, the rows of $\Gamma$ represent a basis for $\mathrm{ker}(\mathscr{H}^\top_L(\bar{w}))$.
\end{proof}

We emphasize that $\bar{w}$ represents some trajectory of the system and \textit{not} available data. As given by Theorem~\ref{thm_identifiability}, if $\mathrm{rank}(\mathscr{H}_L(\bar{w}))=mL+n$, then $\mathscr{H}_L(\bar{w})$ represents a non-parametric model of the restricted behavior of the system $\mathscr{B}|_L$, i.e., im$(\mathscr{H}_L(\bar{w}))=\mathscr{B}|_L$. The following theorem is the main result of this section. It shows how one can obtain an alternative non-parametric model of the system. In particular, we show that
\begin{equation}
	\mathrm{im}(P)=\mathscr{B}|_L,\qquad \textup{where }P=\mathrm{null}(\Gamma).
\end{equation}

This non-parametric representation has two advantages: (i) it can be obtained from a set of irregular measurements $w^-$ of a potentially (very) small number of data points (that depends on $d$ but is independent of $L$) as in Algorithm~\ref{alg_generalalgorithm} and (ii) does not result in an overparameterization of the spanned input-output trajectories (see Corollary~\ref{cor_efficientFL} below). We point out that a conceptually similar result was reported in \cite[eq. ($\mathscr{B}|_L, \textup{KER}$)]{Markovsky22} but no proofs were provided to show this result.

\begin{theorem}\label{thm_equivalentmatrix}
	Consider an LTI system $\mathscr{B}\in\partial\mathscr{L}_{m,n,\ell}^{q}$ and let $\bar{w}\in\mathscr{B}|_{\bar{T}}$ . For $L\geq\ell+1$, let $\mathrm{rank}(\mathscr{H}_L(\bar{w}))=mL+n$ and let the rows of $\Gamma$ denote a basis for its left kernel, i.e., $\Gamma\mathscr{H}_L(\bar{w})=0$. Then a matrix of the form $P=\mathrm{null}(\Gamma)$ has $\mathrm{rank}(P)=mL+n$ and satisfies $\mathrm{im}(P)=\mathrm{im}(\mathscr{H}_L(\bar{w}))$, i.e., $P$ represents a non-parametric model of $\left.\mathscr{B}\right|_L$.
\end{theorem}
\begin{proof}
	Let the matrix $\bar{M}\in\mathbb{R}^{qL\times mL+n}$ be formed by $mL+n$ linearly independent columns of $\mathscr{H}_L(\bar{w})$. Clearly,
	\begin{equation}
		\mathrm{im}(\mathscr{H}_L(\bar{w}))=\mathrm{im}(\bar{M}).\label{eqn_imgMbar}
	\end{equation}
	Note that $mL+n < qL$, which holds since $q=m+p$, $L\geq \ell+1$ by assumption and $p\ell \geq n$. Thus, $\bar{M}$ has a non-trivial left kernel, whose basis is also given by the rows of $\Gamma$, i.e., $\Gamma\bar{M}=0$. Recall that $\Gamma$ is a full row rank matrix since it constitutes a basis for the left kernel of $\mathscr{H}_L(\bar{w})$. Furthermore, it has a non-trivial right kernel of dimension $qL-(pL-n)=mL+n$, i.e., there exists a matrix $P=\mathrm{null}(\Gamma)\in~\mathbb{R}^{qL\times mL+n}$. Notice that the matrix $P$ has full column rank (by definition of a basis) and hence, $\mathrm{rank}(P)=mL+n =\mathrm{rank}(\mathscr{H}_L(\bar{w}))$.
	
	Now, it is left to show that $P$ and $\mathscr{H}_L(\bar{w})$ have the same image. By direct sum of orthogonal subspaces of the matrix $\bar{M}$, the following holds
	\begin{equation}
		\begin{aligned}
			\mathrm{im}(\bar{M})\oplus\mathrm{ker}(\bar{M}^\top) = \mathbb{R}^{qL}\\
			\mathrm{im}(\bar{M})\oplus\mathrm{im}(\Gamma^\top) = \mathbb{R}^{qL},
		\end{aligned}\label{eqn_directsumofsubspacesM}
	\end{equation}
	where the second equality holds by definition of $\Gamma$. This implies that $\mathrm{im}(\bar{M})=\mathrm{im}(\Gamma^\top)^\perp$. Similarly, the following holds for the matrix $\Gamma$
	\begin{equation}
		\begin{aligned}
			\mathrm{im}(\Gamma^\top)\oplus\mathrm{ker}(\Gamma) = \mathbb{R}^{qL}\\
			\mathrm{im}(\Gamma^\top)\oplus\mathrm{im}(P) = \mathbb{R}^{qL},
		\end{aligned}\label{eqn_directsumofsubspacesB}
	\end{equation}
	where the second equality holds by definition of $P$. This implies that $\mathrm{im}(\Gamma^\top)^\perp = \mathrm{im}(P)$. Since $\mathrm{im}(\bar{M})=\mathrm{im}(\Gamma^\top)^\perp = \mathrm{im}(P)$, it follows that $\mathrm{im}(\bar{M}) = \mathrm{im}(P)$. This, together with \eqref{eqn_imgMbar} completes the proof.
\end{proof}

Theorem \ref{thm_equivalentmatrix} illustrates how a non-parametric model of the restricted behavior, other than im$(\mathscr{H}_L(\bar{w}))$, can be obtained as the image of an equivalent matrix $P=\mathrm{null}(\Gamma)$, where the rows of $\Gamma$ represent a basis for $\mathrm{ker}(\mathscr{H}^\top_L(\bar{w}))$. By recalling that $\Gamma$ can be constructed from $R_d$ as in \eqref{Bmatrix}, one can therefore obtain the restricted behavior of $\mathscr{B}|_L$ for any $L\geq d$. It is important to highlight that this allows us to use the available measurements (possibly irregular and of small number of data points) to retrieve the \textit{complete} restricted behavior for \textit{any} length $L$, using three steps: 
\begin{itemize}
	\item[(i)] Using irregularly measured data, compute the kernel representation using, e.g., Algorithm \ref{alg_generalalgorithm}.
	\item[(ii)] Build the matrix $\Gamma$ as in \eqref{Bmatrix}.
	\item[(iii)] Obtain $P=\mathrm{null}(\Gamma)$, which according to Theorem~\ref{thm_equivalentmatrix} satisfies im$(P)=\mathscr{B}|_L$ for $L\geq d$.
\end{itemize}

The following corollary follows from Theorem \ref{thm_equivalentmatrix} and provides an analogous result to that of Corollary~\ref{cor_NonParametricRepresentation} in the case of missing data. In particular, we provide a data-based representation for any complete trajectory of the system $\bar{w}$ as a linear combination of the columns of the matrix $P$ which is constructed by post-processing a sequence of irregularly measured (and potentially short number of) data $w^-$ as in steps (i)-(iii) above.

\begin{corollary}\label{cor_efficientFL}
	Given irregular measurements $w^-\in(\mathbb{R}_{\textup{ext}}^{q})^T$ of a trajectory $w\in\mathscr{B}|_T$ where $\mathscr{B}\in\partial\mathscr{L}_{m,n,\ell}^{q}$, suppose Algorithm~\ref{alg_generalalgorithm} returns a matrix $R_d$ with $\mathrm{rank}(R_d)=pd-n$. Then, $\bar{w}\in\mathscr{B}|_L$ for any $L\geq d$, if and only if there exists a vector $\beta\in\mathbb{R}^{mL+n}$ such that
	\begin{equation}
		P\beta=\bar{w},\label{eqn_efficientFL}
	\end{equation}
	where $P=\mathrm{null}(\Gamma)$ and $\Gamma$ is built from $R_d$ as in \eqref{Bmatrix}.
\end{corollary}

Similar to Corollary~\ref{cor_NonParametricRepresentation} or the fundamental lemma \cite{Willems05}, Corollary~\ref{cor_efficientFL} provides us with a data-based representation of the length-$L$ trajectories of LTI systems (i.e., $\mathscr{B}|_L$) given by the image of the matrix $P$. However, there are two important differences between Corollary~\ref{cor_efficientFL} and Corollary~\ref{cor_NonParametricRepresentation}. (i) Even in the case of complete data, Corollary~\ref{cor_efficientFL} retrieves $\mathscr{B}|_L$ \textit{for any $L\geq d$} using a small number of data points which are independent of $L$, specifically $T\geq (m+1)(d+n)-1$. In contrast, Corollary~\ref{cor_NonParametricRepresentation} (or \cite{Willems05}) requires a number of data points which grows as $L$ grows ($T\geq (m+1)(L+n)-1$). (ii) Unlike the fundamental lemma \cite{Willems05} or Corollary~\ref{cor_NonParametricRepresentation}, the vector $\beta\in\mathbb{R}^{mL+n}$ is \textit{unique} (for every $\bar{w}\in\mathscr{B}|_L$) since $P$ has full column rank. This is expected since, in order to uniquely define a length-$L$ trajectory of the LTI system, one needs $mL$ parameters for the input and $n$ to fix the initial conditions of the state (cf. \cite{Berberich20}). In \cite{alsalti2023eddpc_ecc}, we recently exploited these two important differences to propose a data-driven predictive control scheme which is both sample and computationally more efficient than existing ones in the literature.

Recall that Algorithm~\ref{alg_generalalgorithm} takes as an input an incomplete sequence of measurements $w^-$ of length $T$ and, if successful, it returns the kernel representation of the system. In the literature of identification from missing data, other techniques rely on first completing the missing data sequence and then computing the kernel representation as the basis of the left null space of the completed Hankel matrix. The former step of data completion is solved by, e.g., structured low-rank data completion which is a nonlinear optimization problem (compare \cite{Markovsky13b} and references therein). Convex relaxations of this problem use the nuclear norm heuristic (cf. \cite{Liu13}). However, these methods are not computationally efficient.
\begin{algorithm}[!t]
	\caption{Data completion}\label{alg_matrixcompletion}
	\textbf{Input:} Irregular measurements $w^-\in(\mathbb{R}_{\textup{ext}}^{q})^T$ of a trajectory $w\in\mathscr{B}|_T$ where $\mathscr{B}\in\partial\mathscr{L}_{m,n,\ell}^{q}$ is of known complexity.
	\begin{itemize}
		\item[1.] Run Algorithm~\ref{alg_generalalgorithm} to obtain $R_d$.
		\item[2.] For $L=T$, build the matrix $\Gamma$ as in \eqref{Bmatrix}.
		\item[3.] Obtain $P=\mathrm{null}(\Gamma)$.
		\item[4.] Solve $P_g\beta=w_g$ for $\beta$, then obtain $w^-_{m}=P_{m}\beta$.
	\end{itemize}
	\textbf{Output:} A completed sequence $w$ where $w^-_m$ is inserted in place of NaN values in $w^-$.
\end{algorithm}%

If the data is generated by an LTI system, then Corollary~\ref{cor_efficientFL} offers an alternative method to solve the data completion problem, upon successful completion of Algorithm~\ref{alg_generalalgorithm}. In particular, if we denote the given elements of $w^-$ by $w^-_{g}$ and the missing elements of $w^-$ by $w^-_{m}$, then one can use Corollary~\ref{cor_efficientFL}, with $L=T$, to solve $P_{g}\beta = w^-_{g}$ for $\beta$, where $P_{g}$ denotes the submatrix of $P$ given by the rows whose indices match the row indices of $w^-_g$. Notice that the system of equations $P_{g}\beta = w^-_{g}$ is consistent and, hence, there always exists a solution $\beta$ for any $w^{-}_g$. Once a solution $\beta$ is obtained, one can complete the sequence $w$ by solving for $w_{m}^{-}=P_m\beta$, where $P_m$ denotes the submatrix of $P$ whose row indices match the row indices of $w_m^-$. This approach to data completion is summarized in Algorithm~\ref{alg_matrixcompletion}. In Section~\ref{sec_examples}, we illustrate how Algorithm~\ref{alg_matrixcompletion} is computationally more efficient than the nuclear norm heuristic for data completion. However, it hinges on the successful completion of Algorithm~\ref{alg_generalalgorithm}.
\begin{remark}
	Note that, depending on the pattern of missing data, the outcome of Algorithm~\ref{alg_matrixcompletion} may not be unique. However, the completed sequence is unique if and only if $P_g$ has full column rank, specifically $\mathrm{rank}(P_g)=mT+n$. This latter rank condition depends on the pattern of given data.
\end{remark}
In the following section, we consider the case of periodically missing outputs with period $\ell+1$, and impose conditions on the input and the system under consideration such that a kernel representation is guaranteed to be retrieved using Algorithm~\ref{alg_generalalgorithm}. Combined with the results of Theorem~\ref{thm_equivalentmatrix} and Corollary~\ref{cor_efficientFL} above, it is shown that any \textit{complete} finite-length behavior of the system can be retrieved from a set of measurements with periodically missing outputs.

%% file: sections/periodic_SO.tex
\section{Periodically missing outputs}\label{sec_periodicSO}
In order to guarantee that Algorithm~\ref{alg_generalalgorithm} succeeds in returning the kernel representation, we must impose some assumptions on the pattern of the missing data and provide PE-like conditions on the input to the system. In this section, we assume that the system admits a state-space representation as in \eqref{eqn_SSrepresentation} with $(A,B)$ controllable and a known input-output partitioning of the data $w(t)=\,\begin{bsmallmatrix} u(t) \\ y(t) \end{bsmallmatrix}$. Output data is assumed to be periodically missing with period $\ell+1$. For this pattern of missing data, the sequence $w^-\in(\mathbb{R}_{\textup{ext}}^{q})^T$ takes the form
\begin{equation}
	w^-\hspace{-0.5mm} = \hspace{-0.5mm}\scalebox{0.95}{$\left(w(0),\cdots,w(\ell-1),\begin{psmallmatrix}
			u(\ell)\\\textup{NaN}
		\end{psmallmatrix},w(\ell+1),\cdots,w(T-1)\right)$}.\label{eqn_PMO}
\end{equation}
We further assume that $\mathrm{rank}(C)=p<n$ (i.e., $\ell>1$). This is because when $\ell=1$, every other measurement will be lost and hence, no two consecutive measurements are available to allow for identification of the system dynamics.

The considered pattern of periodically missing outputs with period $\ell+1$ is interesting to investigate since a missing data point will appear in every row and every column of the Hankel matrix $\mathscr{H}_d(w^-)$ for $d\geq \ell+1$. In contrast, for data which is periodically missing with period greater than $\ell+1$, one can find conditions on $T$ such that the Hankel matrix $\mathscr{H}_{\ell+1}(w^{-})$ has $m(\ell+1)+n$ complete columns. If these columns are linearly independent, then a kernel representation can directly be obtained by taking the left kernel of the submatrix containing the aforementioned linearly independent columns. For data missing with periodicity less than $\ell+1$, Algorithm~\ref{alg_generalalgorithm} does not return a kernel representation of the system. In \cite{markovsky24} the so-called lifting operator is employed to facilitate identification of the dynamics on a coarser time scale and later recovers the true system by the reversing the effect of the lift operator. Such a method can handle periodically missing outputs with period less than $\ell+1$ (including $\ell=1$), but requires further (more restrictive) assumptions on the system to that end.

We start by illustrating how to systematically choose the submatrices $\bar{H}_i$ in Algorithm \ref{alg_generalalgorithm}. Later, we provide conditions on the input to the system such that certain submatrices satisfy a desired rank (a PE-like condition), such that Algorithm \ref{alg_generalalgorithm} is guaranteed to return the kernel representation of the system. 

Consider a sequence of irregular measurements $w^-\in(\mathbb{R}_{\textup{ext}}^{q})^T$ where the output portion is periodically missing with period equal to $\ell+1$. The following two matrices are submatrices of the Hankel matrix $\mathscr{H}_{\ell+2}(w^-)$ (i.e., $d=\ell+2$)
\begin{equation}
\begin{aligned}
	&\hspace{-2mm}\scalebox{1}{${H}_1\hspace{-1mm}=\hspace{-1mm}\begingroup
		\setlength\arraycolsep{1.25pt}\begin{bmatrix}
			w(0) & w({\ell+1}) & \cdots & w({(n_{c}-1)(\ell+1)})\\
			\vdots & \vdots & \vdots & \vdots\\
			w({\ell-1}) & w({2\ell}) & \cdots & w({(n_{c}-1)(\ell+1)+\ell-1})\\
			\begin{pmatrix}
				u({\ell})\\\textup{NaN}
			\end{pmatrix} & \begin{pmatrix}
				u({2\ell+1})\\\textup{NaN}
			\end{pmatrix} & \cdots & \begin{pmatrix}
				u({(n_{c}-1)(\ell+1)+\ell})\\\textup{NaN}
			\end{pmatrix}\\
			w({\ell+1}) & w({2\ell+2}) & \cdots & w({n_{c}(\ell+1)})
		\end{bmatrix}\endgroup\hspace{-1mm},$}\\
	&\hspace{-2mm}\scalebox{1}{${H}_2\hspace{-1mm}=\hspace{-1mm}\begingroup
			\setlength\arraycolsep{1.25pt}\begin{bmatrix}
				w(1) & w({\ell+2}) & \cdots & w({(n_{c}-1)(\ell+1)+1})\\
				\vdots & \vdots & \vdots & \vdots\\
				\begin{pmatrix}
					u({\ell})\\\textup{NaN}
				\end{pmatrix} & \begin{pmatrix}
					u({2\ell+1})\\\textup{NaN}
				\end{pmatrix} & \cdots & \begin{pmatrix}
					u({(n_{c}-1)(\ell+1)+\ell})\\\textup{NaN}
				\end{pmatrix}\\
				w({\ell+1}) & w({2\ell+2}) & \cdots & w({n_{c}(\ell+1)})\\
				w(\ell+2) & w({2\ell+3}) & \cdots & w({n_{c}(\ell+1)+1})
			\end{bmatrix}\endgroup\hspace{-1mm}.$}
\end{aligned}
\label{eqn_submatricesH1H2_SO}
\end{equation}

After deleting the NaN block rows, we are left with two matrices $\bar{H}_1,\bar{H}_2\in\mathbb{R}^{q(\ell+2)-p\times n_{c}}$, where $n_c$ denotes the number of columns. Lemma~\ref{lemma_PElikeCondition} below provides conditions on the input such that $\mathrm{rank}(\bar{H}_1)=\mathrm{rank}(\bar{H}_2)=m(\ell+2)+n$. This rank condition is needed for the submatrices to uncover part of the kernel representation of the system (compare Lemma~\ref{lemma_submatrices} and Algorithm~\ref{alg_generalalgorithm}). But first, we make the following assumption which is satisfied, e.g., for systems whose matrix $A$ has only positive real eigenvalues (cf. \cite{Krauss22}).

\begin{assumption}\label{assmp_SBO}
	For a system $\mathscr{B}\in\partial\mathscr{L}_{m,n,\ell}^{q}$ with $\mathscr{B}_{ss}$ as in \eqref{eqn_SSrepresentation}, the following matrix has full rank
	\begin{equation*}
		\mathscr{O}_s = \begingroup
		\setlength\arraycolsep{1.5pt}\begin{bmatrix}
			C^\top & (CA)^\top & \cdots & (CA^{\ell-2})^\top & (CA^{\ell+1})^\top
		\end{bmatrix}\endgroup^\top\in\mathbb{R}^{p\ell \times n}.\label{sample_based_obsv}
	\end{equation*}
\end{assumption}

\begin{lemma}\label{lemma_PElikeCondition}
	Let Assumption~\ref{assmp_SBO} hold and let the input sequence applied to a controllable system $\mathscr{B}\in\partial\mathscr{L}_{m,n,\ell}^{q}$ satisfy $\mathrm{rank}\left(\mathcal{H}_{[0,n+\ell+2]}(u)\right) = m(n+\ell+3)$ where
	\begin{align}
		&\mathcal{H}_{[0,n+\ell+2]}(u) \coloneqq\label{eqn_MosaicPEinputs}\\
		&\begingroup
		\setlength\arraycolsep{2pt}\begin{bmatrix}
			u_{[0,n+\ell+2]} &\sigma^{\ell+1}u_{[0,n+\ell+2]} & \ldots & \sigma^{(n_c-1)(\ell+1)}u_{[0,n+\ell+2]}
		\end{bmatrix}\endgroup.\notag
	\end{align}
	Then the corresponding resulting matrices $\bar{H}_1$ and $\bar{H}_2$ in \eqref{eqn_submatricesH1H2_SO} have $\mathrm{rank}(\bar{H}_1)=\mathrm{rank}(\bar{H}_2)=m(\ell+2)+n$.
\end{lemma}
\begin{proof}
	We will prove the claim for $\bar{H}_1$. Similar steps can be made to show the claim for $\bar{H}_2$. Following a row permutation, the matrix $\bar{H}_1$ can be equivalently written as
	\begin{equation*}
		\scalebox{0.9}{$\begin{aligned}
				\bar{H}_1 &= \bar{\Pi}_r\begingroup
				\setlength\arraycolsep{1.5pt}\begin{bmatrix}
					\mathcal{H}_{[0,\ell+1]}(u) \\ \hline 
					\begin{matrix}
						y(0) & y({\ell+1}) & \cdots & y({(n_{c}-1)(\ell+1)})\\
						\vdots & \vdots & \vdots & \vdots\\
						y({\ell-1}) & y({2\ell}) & \cdots & y({(n_{c}-1)(\ell+1)+\ell-1})\\
						y({\ell+1}) & y({2\ell+2}) & \cdots & y({n_{c}(\ell+1)})
					\end{matrix}
				\end{bmatrix}\endgroup,
			\end{aligned}$}
	\end{equation*}
	where $\bar\Pi_r$ is a square full rank permutation matrix of appropriate dimensions. Considering some minimal state-space representation of the system, this can be written as
	\begin{align*}
		\bar{H}_1 \scalebox{0.9}{$=\bar{\Pi}_r \underbrace{\left[\begin{array}{c|c}
					I & {0}\\ \hline
					\begingroup
					\setlength\arraycolsep{2pt}\begin{matrix}
						D & 0 & \cdots & 0\\
						\vdots & \vdots & \vdots & \vdots\\
						CA^{\ell-2}B & CA^{\ell-3}B & \cdots & 0\\
						CA^{\ell}B & CA^{\ell-1}B & \cdots & D\\
					\end{matrix}\endgroup & \begin{matrix}
						C\\ \vdots \\ CA^{\ell-1}\\ CA^{\ell+1}
					\end{matrix}
				\end{array}\right]}_{\coloneqq \mathscr{T}_1\in\mathbb{R}^{m(\ell+2)+p(\ell+1) \times m(\ell+2)+n}}\begin{bmatrix}
				\mathcal{H}_{[0,\ell+1]}(u) \\ X_0
			\end{bmatrix},$}
	\end{align*}
where $X_0\coloneqq \begin{bmatrix}
	x(0) & x(\ell+1) & \cdots & x((n_{c}-1)(\ell+1))
\end{bmatrix}$. The matrix $\mathscr{T}_1$ has full column rank. This is due to the identity on the upper left block and the first $\ell$ blocks of the observability matrix on the right lower block\footnote{When showing the claim for $\bar{H}_2$, the bottom right block of $\mathscr{T}_2$ contains instead a sample-based observability matrix which is assumed to have full column rank by Assumption~\ref{assmp_SBO}.}. Hence, it holds that $\mathrm{rank}(\bar{H}_1)=\mathrm{rank}\left(\begin{bsmallmatrix}\mathcal{H}_{[0,\ell+1]}(u) \\ X_0\end{bsmallmatrix}\right)$. Now, it is left to show that $\mathrm{rank}\left(\begin{bsmallmatrix}\mathcal{H}_{[0,\ell+1]}(u) \\ X_0\end{bsmallmatrix}\right) = m(\ell+2)+n$. Notice that the deeper matrix $\mathcal{H}_{[0,n+\ell+2]}(u)$ has rank $m(n+\ell+3)$ by assumption, which implies that $\mathcal{H}_{[0,n+\ell+1]}(u)$ has rank $m(n+\ell+2)$. This means that the input sequences given by the columns of $\mathcal{H}_{[0,n+\ell+1]}(u)$ are collectively persistently exciting (see \cite[Def. 2]{vanWaarde20}) of order $n+\ell+2$. Since the system is controllable, it follows from \cite[Thm. 2.i]{vanWaarde20} that $\mathrm{rank}\left(\begin{bsmallmatrix}\mathcal{H}_{[0,\ell+1]}(u) \\ X_0\end{bsmallmatrix}\right) = m(\ell+2)+n$.
\end{proof}

\begin{remark}
	Notice that Lemma~\ref{lemma_PElikeCondition} necessitates that $n_{c}\geq m(\ell+n+3)$ and hence the length of $w^-$ must be at least $T\geq m(\ell+n+3)(\ell+1)+n+2$.
\end{remark}

When the submatrices $\bar{H}_1,\bar{H}_2$ have the desired ranks as in the above lemma, their left kernels (the bases of which are given by full row rank matrices $\bar{Z}_i$ satisfying $\bar{Z}_i\bar{H}_i=0$) have the following dimensions
\begin{align*}
	\mathrm{dim(ker}(\bar{H}_i^\top))&=\#\mathrm{rows}(\bar{H}_i) - \mathrm{rank}(\bar{H}_i)\\
	&= q(\ell+2)-p - m(\ell+2)-n\\
	&= p(\ell+1)-n.
\end{align*}

By inserting zero columns in the location of deleted block rows of the corresponding $H_i$, one can extend $\bar{Z}_i$ to $Z_i$ which then satisfies $Z_i\mathscr{H}_{\ell+2}(w)=0$, for $i={1,2}$ (compare Example~\ref{example_MotivatingExample} and Lemma~\ref{lemma_submatrices}). Concatenating the two, we get
\begin{equation*}
	\begin{bsmallmatrix}
		Z_1\\ Z_2
	\end{bsmallmatrix}\mathscr{H}_{\ell+2}(w)=\mathscr{Z}\mathscr{H}_{\ell+2}(w)=0,
\end{equation*}
where $\mathscr{Z}\in\mathbb{R}^{2p(\ell+1)-2n \times q(\ell+2)}$. The following theorem shows that $\mathscr{Z}$ contains $p(\ell+2)-n$ linearly independent rows that specify a kernel representation of the system.
\begin{theorem}\label{thm_rankZ}
	Consider a controllable system $\mathscr{B}\in\partial\mathscr{L}_{m,n,\ell}^{q}$ as in \eqref{eqn_SSrepresentation} with $\mathrm{rank}(C)<n$ and let Assumption~\ref{assmp_SBO} hold. Given irregular measurements $w^{-}\in(\mathbb{R}^{q}_{\textup{ext}})^T$ of the form \eqref{eqn_PMO} of a trajectory $w\in\mathscr{B}|_{T}$, let $\mathrm{rank}(\mathcal{H}_{[0,n+\ell+2]}(u))=m(n+\ell+3)$ where $\mathcal{H}_{[0,n+\ell+2]}(u)$ is as in \eqref{eqn_MosaicPEinputs}. Then, Algorithm~\ref{alg_generalalgorithm} successfully returns a matrix $R_{\ell+2}$ which specifies a kernel representation of the system and is given by $p(\ell+2)-n$ linearly independent rows of the matrix $\mathscr{Z}$.
\end{theorem}
\begin{proof}
	See Appendix~\ref{appndx_proofThmZ}.
\end{proof}

Combining the results of Corollary~\ref{cor_efficientFL} and Theorem~\ref{thm_rankZ}, we obtain the following corollary, which is analogous to \cite[Thm. 1]{Willems05} for the case of periodically missing data. In particular, we provide conditions on the input such that any \textit{complete} finite-length behavior of the system can be obtained. Unlike \cite{Willems05}, we only have periodically missing data and the resulting representation of all input/output trajectories is not overparametrized (see the discussion following Corollary~\ref{cor_efficientFL}).

\begin{corollary}
	Consider a controllable system $\mathscr{B}\in\partial\mathscr{L}_{m,n,\ell}^{q}$ and let the conditions of Theorem~\ref{thm_rankZ} hold. Then, for $L\geq \ell+2$, $\bar{w}\in\mathscr{B}|_L$ if and only if there exists $\beta\in\mathbb{R}^{mL+n}$ such that
	\begin{equation}
		P\beta=\bar{w},
	\end{equation}
	where $P=\mathrm{null}(\Gamma)$ and $\Gamma$ is given as in \eqref{Bmatrix}.
\end{corollary}

So far, we only considered exact (i.e., noise-free) data. When the data is noisy, then the rank condition in \eqref{rank_condition_L} is, in general, violated. This necessitates certain modifications of our results in Algorithm~\ref{alg_generalalgorithm}. In the following section, we consider the case of noisy and irregularly measured data, and discuss the effect of noise on the results of Algorithm~\ref{alg_generalalgorithm}.

%% file: sections/noisydata.tex
\section{Handling noisy data}\label{sec_noisydata}
In this section, we consider (irregular) measurements which are affected by noise. In particular, one has access to $\widetilde{w}^-$ whose (available) elements take the form $\widetilde w_k = w_k + \varepsilon_k$ (errors-in-variables setting), where $\varepsilon$ denotes some noise realization, e.g., sampled from a Gaussian or uniform distribution. It is important to note that, even when the available data is complete, the presence of noise, in general, results in the violation of the rank condition \eqref{rank_condition_L}. In fact, additive uniform random noise or Gaussian noise will almost surely increase the rank i.e., $\mathrm{rank}(\mathscr{H}_L(\widetilde w))\geq mL+n$. Such a rank condition, however, is crucial for obtaining data-based representations of LTI systems (cf. Theorem~\ref{thm_identifiability}). In this case, a common practice is to obtain a low rank approximation $\widehat{\mathscr{H}}_L(\widetilde w)$ of $\mathscr{H}_L(\widetilde w)$ with $\mathrm{rank}(\widehat{\mathscr{H}}_L(\widetilde w)) = mL+n$ and then use it as an estimate of the true finite-length behavior of the LTI system that generated the data, i.e., $\widehat{\mathscr{B}}|_{L}\coloneqq \mathrm{im}(\widehat{\mathscr{H}}_L(\widetilde w))$ (cf. \cite{Markovsky08}). Low rank approximation can be done using, e.g., truncated singular-value decomposition (TSVD, cf. \cite{Eckart36}), or using structured low-rank approximation of Hankel matrices (SLRA, cf. \cite{Markovsky08}).

In Section~\ref{sec_algorithm}, we addressed the noise-free setting by proposing Algorithm~\ref{alg_generalalgorithm} which, if successful, returns a basis for the left kernel of a Hankel matrix with missing entries. An important step in Algorithm~\ref{alg_generalalgorithm} involved selecting submatrices (denoted $\bar{H}_i$) with specific rank. Since, in general, the presence of noise will result in full rank submatrices $\bar{H}_i$ in Algorithm~\ref{alg_generalalgorithm}, it is not clear anymore how to proceed in order to retrieve a basis for the left kernel. To that end, we propose modifying Algorithm~\ref{alg_generalalgorithm} by introducing an intermediate low rank approximation step. In particular, one would now search over submatrices $\bar{H}_i$ with $\mathrm{rank}(\bar{H}_i)\geq md+n$ then apply TSVD to obtain an approximate matrix, denoted $\widehat{H}_i$, such that $\mathrm{rank}(\widehat{H}_i)=md+n$. The rest of the procedure continues as in Algorithm~\ref{alg_generalalgorithm} with the main difference that the exit condition on the rank of $\mathscr{Z}$ is an inequality (rather than an equality). This is because $\mathscr{Z}$ now contains vectors that do not necessarily represent annihilators and, hence, $\mathscr{Z}$ can have larger rank (i.e., corresponding to a system of different complexity than the true system). To overcome this, we introduce another low rank approximation step on $\mathscr{Z}$ and obtain a matrix $\widehat{\mathscr{Z}}$ such that $\mathrm{rank}(\widehat{\mathscr{Z}})=pd-n$. As a result, Algorithm~\ref{alg_noisygeneralalg}, if successful\footnote{Similar to the discussion below Algorithm~\ref{alg_generalalgorithm}, Algorithm~\ref{alg_noisygeneralalg} is deemed successful if it terminates with $\mathrm{rank}(\widehat{\mathscr{Z}})=pd-n$.}, returns a matrix $\widehat{R}_d$ (given by $pd-n$ linearly independent rows of $\widehat{\mathscr{Z}}$) which specifies an \textit{estimate} of the true kernel representation specified by (the unknown) $R_d$. The modification of Algorithm~\ref{alg_generalalgorithm} to account for noisy data is summarized in Algorithm~\ref{alg_noisygeneralalg}.
\begin{algorithm}[!t]
	\caption{Computing a kernel representation using noisy data}\label{alg_noisygeneralalg}
	\textbf{Input:} Noisy and irregular measurements $\widetilde{w}^-\in(\mathbb{R}_{\textup{ext}}^{q})^T$ of a trajectory $w\in\mathscr{B}|_T$ where $\mathscr{B}\in\partial\mathscr{L}_{m,n,\ell}^{q}$ is of known complexity.
	\begin{itemize}
		\item[1.] Construct the extended sequence $\widetilde{w}_{\textup{ext}}$ by appending $\widetilde{w}^{-}$ with $T$ instances of NaN.
		\item[2.] Construct the matrix $\mathscr{H}_T(\widetilde{w}_{\textup{ext}})$, and initialize $\mathscr{Z}=[\,]$.
		\item[3.] \textbf{For} $\ell+1\leq d\leq T$ \textbf{do}
		\begin{itemize}
			\item Search over all submatrices of $\mathscr{H}_d(\widetilde{w}^{-})$, i.e., $\bar{H}_i\in\mathbb{R}^{\eta_i\times \rho_i}$ that have $\mathrm{rank}(\bar{H}_i)\geq md+n$.
			\item Compute a low rank approximation of $\bar{H}_i$, denoted $\widehat{H}_i$, such that $\mathrm{rank}(\widehat{H}_i)=md+n$.
			\item Compute a basis for $\mathrm{ker}(\widehat{H}_i^\top)$, i.e., a full row rank matrix $\widehat{Z}_i\in\mathbb{R}^{\zeta_i\times\eta_i}$ such that $\widehat{Z}_i \widehat{H}_i=0$.
			\item Extend $\widehat{Z}_i$ to $Z_i$ by inserting zero columns at the location of the rows removed from $\mathscr{H}_d(\widetilde{w}^-)$ to obtain $\widehat{H}_i$.
			\item Concatenate all the above matrices $Z_i$ in $\mathscr{Z}$, i.e., $\mathscr{Z}\coloneqq\begin{bmatrix}
				[\begin{matrix}
					\mathscr{Z} & \mathbf{0}
				\end{matrix}]^\top & Z_1^\top & Z_2^\top & \cdots
			\end{bmatrix}^\top$, where $\mathbf{0}$ denotes a matrix of zeros of appropriate dimensions.
			\item \textbf{If} $\mathrm{rank}(\mathscr{Z})\geq pd-n$, \textbf{break}
		\end{itemize}
		\textbf{end for}
		\item[4.] \textbf{If} $\mathrm{rank}(\mathscr{Z}) = pd-n$, \textbf{set} $\widehat{\mathscr{Z}} = \mathscr{Z}$.\\
		\textbf{Else}, compute a low rank approximation of $\mathscr{Z}$, denoted $\widehat{\mathscr{Z}}$, such that $\mathrm{rank}(\widehat{\mathscr{Z}})=pd-n$.
	\end{itemize}
	\textbf{Output:} The matrix $\widehat{R}_d$, specifying an approximate kernel representation of the system, is given by any $pd-n$ linearly independent rows of $\widehat{\mathscr{Z}}$.
\end{algorithm}%

We conclude this section with some remarks on uncertainty quantification in the behavioral framework (see, e.g., \cite{Padoan22,Fazzi23}). A good measure for quantifying the quality of the estimate $\mathrm{ker}(\widehat{R}_d)\eqqcolon \widehat{\mathscr{B}}|_d$ is given by the principal angles \cite{Miao1992} between the true and estimated subspaces, i.e., ${\mathscr{B}}|_d$ and $\widehat{\mathscr{B}}|_d$, respectively. Specifically, two subspaces are said to be close to each other if the largest principle angle $\theta_{\textup{max}}(\mathscr{B}|_d, \widehat{\mathscr{B}}|_d)$, which depends on the noise, is small. In the last step of Algorithm~\ref{alg_noisygeneralalg}, we introduced a low-rank approximation step on the matrix $\mathscr{Z}$ which ensures that the estimated kernel representation corresponds to a system with the same complexity as that of the true system. Therefore, it directly follows that having $\theta_{\textup{max}}(\mathscr{B}|_d,\widehat{\mathscr{B}}|_d)=0$ is equivalent to $\mathscr{B}|_d=\widehat{\mathscr{B}}|_d$. Our future work will focus on providing quantitative error bounds on the angle $\theta_{\textup{max}}(\mathscr{B}|_d, \widehat{\mathscr{B}}|_d)$ in terms of the noise bound.

In the next section, we illustrate the results of this paper with different simulation examples.

%% file: sections/examples.tex
\section{Simulation examples}\label{sec_examples}
In this section, we will first empirically investigate the success/failure of Algorithm~\ref{alg_generalalgorithm} in the case of random patterns of missing input and/or output instances. Then, we compare the computational efficiency of solving a data completion problem using the results of this paper (see discussion after Corollary~\ref{cor_efficientFL}) against the nuclear norm heuristic \cite{Liu13}. Finally, we illustrate the results of Algorithm~\ref{alg_noisygeneralalg} on a (simplified) physiological model of cardiovascular hemodynamics that describes the effect of fluid infusion on the change of blood volume (cf. \cite{bighamian16}). All simulations were done on a standard Intel i7-10875H (2.30 GHz) machine with 16 GB of memory.

\subsection{Performance of Algorithm~\ref{alg_generalalgorithm} under random patterns of missing data}\label{sec_ExA}
In this section, we empirically evaluate the performance of Algorithm~\ref{alg_generalalgorithm} when the data is missing randomly. We generate 100 random discrete-time LTI systems with $(m,p,n,\ell) = (0,1,2,2)$. For each system, we generate a trajectory $w$ of length $T_{\mathrm{max}}=200$ by simulating the response of each system to random initial conditions sampled from a uniform distribution $(U(0,1))^2$. To investigate the performance of Algorithm~\ref{alg_generalalgorithm} in identifying each system, we perform several experiments using irregular measurements $w^{-}\in(\mathbb{R}_{\textup{ext}}^q)^T$ of the above trajectory $w$. For each experiment, we vary the length of $w^{-}\in(\mathbb{R}_{\textup{ext}}^q)^T$ that is passed to the algorithm from $T=20$ to $T=T_{\mathrm{max}}=200$ as well as the percentage of given to missing data from 10\% up to 100\%.
\begin{figure}[!t]
	\begin{center}
		\includegraphics[width=0.75\columnwidth]{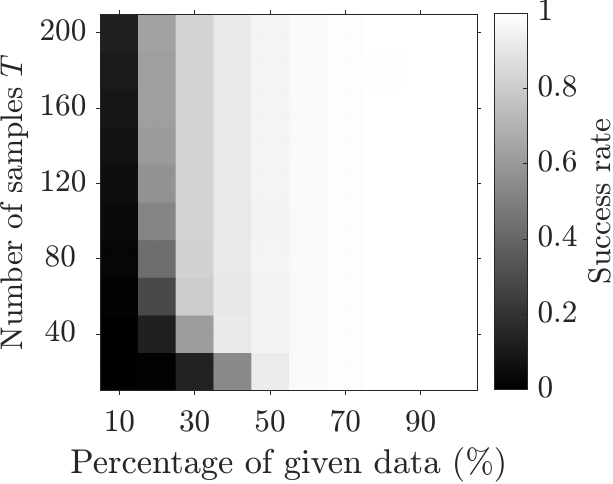}
	\end{center}
	\caption{Transition diagram from failure (black) to success (white) when the data is missing randomly. It can be seen that the success rate of Algorithm~\ref{alg_generalalgorithm} increases as both the length of the sequence $T$ and as the fraction of given to missing samples increase.}
	\label{fig1}
\end{figure}

For each system and percentage of missing data, the following was repeated $100$ times: (i) the location of the missing samples was randomly generated using the MATLAB function \texttt{randperm}, then (ii) Algorithm~\ref{alg_generalalgorithm} was run and checked for success (assigned a value of one) or failure (assigned a value of zero). We additionally validated successful outcomes of Algorithm~\ref{alg_generalalgorithm} by comparing the output of the algorithm to the kernel representation computed from the Hankel matrix of the complete data of the same depth. The average time it took to complete one run of Algorithm~\ref{alg_generalalgorithm} was $0.0381$ seconds. Finally, for each percentage of missing data, the results were averaged over all $100$ runs and all $100$ systems. Figure~\ref{fig1} shows a gray-scale plot of the results. It can be seen how a transition occurs from failure to success (from dark to light) as the length of the sequence $T$ and the fraction of given samples increase. Notice that the algorithm consistently succeeded when less than 30\% of the data are missing. One possible explanation is that the random pattern of missing samples (which was obtained using Matlab's \texttt{randperm} command, see above) did not result in pathological cases where the algorithm fails to find a kernel representation.

\subsection{Data completion: comparison vs nuclear norm heuristic}\label{sec_exB}
Following Corollary~\ref{cor_efficientFL}, we briefly discussed how one can use the results of this paper to solve the data completion problem (cf. Algorithm~\ref{alg_matrixcompletion}). The goal of this example is to compare the computational performance of our proposed method in Algorithm~\ref{alg_matrixcompletion} against the nuclear norm heuristic \cite{Liu13} when solving the data completion problem.

For the test system, we use the one in \cite{Markovsky13} which is a single-output six-dimensional slightly damped autonomous system. A trajectory $w$ of length $T=500$ was generated by simulating the response of the system when all states are initialized from one. For this example, a total of $276$ samples were periodically missing (with different periodicities). Table~\ref{table_ex2} illustrates the results of the data completion problem. The error is measured in terms of the $2-$norm, i.e., $e = \norm*{w-\hat{w}}_2/\norm*{w}_2$, where $\hat{w}$ is the sequence returned by the two methods. When only the first $200$ samples of $w^-$ are used, both approaches manage to solve the data completion problem up to numerical accuracy (in this case, the ratio of missing data is around $33\%$). However it is clear that our proposed method in Algorithm~\ref{alg_matrixcompletion} is much faster than the nuclear norm minimization approach. When using the entire sequence of length $T=500$, the nuclear norm approach failed to return a result (due to insufficient memory). This is caused by the increasingly growing computations that are required to solve an optimization problem (here, we used the default solver in CVX \cite{cvx}). In contrast, Algorithm~\ref{alg_matrixcompletion} only uses linear algebra operations which solve the data completion problem with (fast) computation time even when the entire sequence is used.
\subsection{Case study: physiological systems}\label{sec_ex_physiologicalsys}
In this example, we consider a (simplified LTI) physiological model of cardiovascular hemodynamics that describes the effect of fluid infusion on the change of blood volume (cf. \cite{bighamian16}). In continuous time, this takes the form
\begin{align}
	\frac{d}{dt}\begin{bmatrix}
		\Delta V_B(t)\\ \breve{u}(t)
	\end{bmatrix} &= \begin{bmatrix}
		-K & \frac{K}{1+\alpha}\\
		0 & 0
	\end{bmatrix}\begin{bmatrix}
		\Delta V_B(t)\\ \breve{u}(t)
	\end{bmatrix}+\begin{bmatrix}
		1\\ \frac{1}{1+\alpha}
	\end{bmatrix}u(t)\notag\\
	y(t) &= \begin{bmatrix}
		1 & 0
	\end{bmatrix}\begin{bmatrix}
		\Delta V_B(t)\\ \breve{u}(t)
	\end{bmatrix}\label{eqn_ex_system}
\end{align}
where $\Delta V_B(t)$ is the change in blood volume, $u(t), \breve{u}(t)$ is the rate of fluid infusion and the total infused volume, respectively. The parameter $\alpha$ specifies the steady-state split of fluid gain between the intravascular and extravascular compartments, while the parameter $K$ specifies the inter-compartmental rate of fluid shift. For details, the reader is referred to \cite{bighamian16}. Even in critical care settings, the (change of) blood volume is typically difficult to measure consecutively and, instead, is estimated based on blood hematocrit levels \cite{causey2011validation, ewaldsson2005kinetics}. Furthermore, the patient-specific parameters $\alpha,K$ are unknown and differ from one patient to another. Therefore, it is convenient in such a setting that one applies our proposed data-driven methods, which are suited for handling missing data, in order to recover the finite-length behavior of such unknown dynamical systems.
\begin{table}[!t]
	\caption{Comparison: data completion problem}
	\vspace{-1.5em}
	\begin{center}
		\begin{tabular}{ |l|c|c| }
			\hline
			Compared values & Algorithm~\ref{alg_matrixcompletion} & Nuclear norm \\ \hline
			\begin{tabular}{c|c} $T=200$ & \begin{tabular}{c} Time (s) \\ Error \end{tabular} \end{tabular} & $\begin{matrix}
				0.0932\\
				6.3278\times10^{-14}
			\end{matrix}$ & $\begin{matrix}
				57.9953\\
				9.3051\times10^{-9}
			\end{matrix}$ \\
			\hline
			\begin{tabular}{c|c} $T=500$ & \begin{tabular}{c} Time (s) \\ Error \end{tabular} \end{tabular} & $\begin{matrix}
				0.2258\\
				6.1301\times10^{-14}
			\end{matrix}$ & $\begin{matrix}
				\textup{N/A}\\
				\textup{N/A}
			\end{matrix}$ \\
			\hline
		\end{tabular}
		\label{table_ex2}
	\end{center}
\end{table}

\begin{figure}[!t]
	\begin{center}
		\includegraphics[width=\columnwidth]{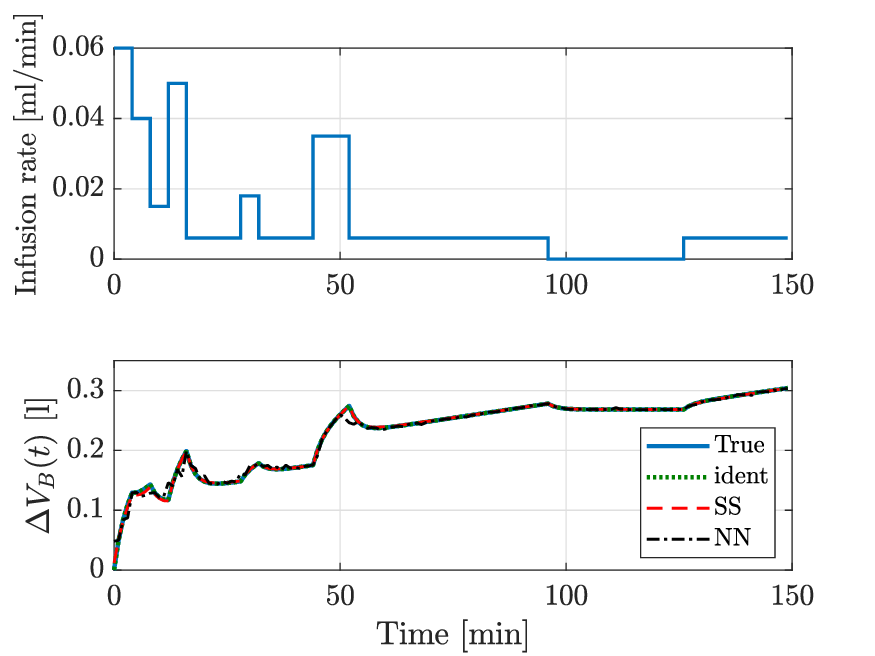}
	\end{center}
	\caption{An illustration of the true vs. estimated trajectories using the three methods: structured low-rank approximation method (\texttt{ident}), our proposed subspace method (SS) and the nuclear norm method (NN). Unlike \texttt{ident}, SS does not require prior model knowledge or initialization steps but still performs similarly to \texttt{ident}.}
	\label{fig_ExC}
\end{figure}

This example aims to illustrate the usefulness of our results when blood volume measurements are only irregularly measured and are further affected by measurement noise. Specifically, we will use Algorithm~\ref{alg_noisygeneralalg} to retrieve an approximate kernel representation of the system, which is later used to retrieve an estimate of the complete finite length behavior of the system as well as an estimate of the missing samples. After discretizing the model \eqref{eqn_ex_system} with a sampling time of $T_s = 1~\mathrm{min}$, we run an open loop simulation where fluid is administered in boluses (i.e., not random inputs, see Figure~\ref{fig_ExC}), and collect corresponding input-output data of length $T=150~\mathrm{min}$. The (unknown) parameters used are $K=0.5, \alpha=1.3$. We consider the case where $40\%$ of the output samples are (randomly) missing, while the remaining samples are affected by noise $\varepsilon\in\mathbb{R}^{pT}$ sampled from a zero-mean Gaussian distribution scaled to different noise levels, i.e., $\widetilde{y}=y+\gamma \frac{\norm{y}}{\norm{\varepsilon}}\varepsilon$ for $\gamma\in\{0,2\times 10^{-4},4\times 10^{-4},6\times 10^{-4},8\times 10^{-4},10^{-3}\}$.

For each value of $\gamma$, we run Algorithm~\ref{alg_noisygeneralalg} to retrieve an approximate kernel representation of the system $\widehat{R}_d$, which is later used to retrieve an estimate of the complete finite length behavior of the system of length equal to $T$ (see steps (i)-(iii) before Corollary~\ref{cor_efficientFL}). This is later used to estimate the missing output samples in the original sequence (using similar steps as in Algorithm~\ref{alg_matrixcompletion}). We refer to this as the subspace method (SS) and compare its performance against the nuclear norm heuristic (NN). Another method to solve the approximate data completion problem is using the \texttt{ident} method for identification in the behavioral setting \cite{ident} which is based on structured low-rank approximation \cite{Markovsky08}. This latter method requires prior model knowledge to initialize the algorithm, which we set to be the exact true model.

Table~\ref{table_exC} contains a summary of the estimation error measured in terms of the $2-$norm, i.e., $e = \norm*{w-\hat{w}}_2/\norm*{w}_2$, where $\hat{w}$ is the sequence returned by the three methods. It is important to point out that the \texttt{ident} method computes the maximum likelihood estimator for the true behavior (see \cite[Prop. 6]{Markovsky08}). Since, moreover, it is initialized at the true model, \texttt{ident} returns the theoretically best possible result using the available data and, hence, is viewed as the benchmark against which SS and NN methods are compared. It can be seen that our direct method (SS) returns very close results to that of \texttt{ident}, without using any prior knowledge about the system. This illustrates the advantages of Algorithm~\ref{alg_noisygeneralalg} which does not use any prior model knowledge (except known complexity), but still manages to return good results compared to \texttt{ident}. For the nuclear norm method (NN), we notice the following: (i) NN performs worse in presence of noise, and fails to complete the sequence even when there is no noise in the data. This is not too surprising as the nuclear norm method is a heuristic convex relaxation of the exact (nonlinear) data completion problem. (ii) Although NN performs worse than \texttt{ident} and SS, it is more robust (i.e., errors do not vary depending on noise level). This is potentially due to the fact that NN does not impose any structure/rank constraint on the matrix and returns the best fit of the data by minimizing the nuclear norm of the Hankel matrix. Since the noise is sampled from a zero mean distribution, small levels of noise $\gamma$ did not largely affect the resulting fitting errors. Figure~\ref{fig_ExC} shows a representative result for the true vs. estimated trajectory using the three different methods when $\gamma=10^{-3}$.

\begin{table}[!t]
	\caption{Comparison: approximate data completion.}
	\vspace{-1.5em}
	\begin{center}
		\begin{tabular}{ |l|c|c|c| }
			\hline
			$\gamma$ & \texttt{ident} & SS & NN\\ \hline
			$0$ & 0 & 0 & 0.0284 \\ \hline
			$2\times 10^{-4}$ & 0.0001 & 0.0014 & 0.0284 \\ \hline
			$4\times 10^{-4}$ & 0.0001 & 0.0046 & 0.0284 \\ \hline
			$6\times 10^{-4}$ & 0.0002 & 0.0041 & 0.0286 \\ \hline
			$8\times 10^{-4}$ & 0.0003 & 0.0018 & 0.0284 \\ \hline
			$10^{-3}$ & 0.0003 & 0.0072 & 0.0283 \\ \hline
		\end{tabular}
		\label{table_exC}
	\end{center}
\end{table}

%% file: sections/conclusion.tex
\section{Discussion and Conclusions}\label{sec_conclusion}
In this paper, we investigated the problem of obtaining non-parametric data-based system representations from a set of irregularly measured data. For the general case of multi-input multi-output systems and random patterns of missing data, we improved an existing algorithm from the literature which now allows for computing the (not necessarily minimal) kernel representation of the system from irregularly measured data. Assuming that the system's complexity is known, one can check if the output of the algorithm corresponds to the kernel representation upon running the algorithm. 

If successful, we showed how one can use the output of the algorithm to obtain any \textit{complete} finite-length behavior of the system by exploiting the properties of the Hankel matrix structure. This result has important properties, namely that the complete behavior of any length can be retrieved independent of the number of available (irregularly measured) values. For the case of periodically missing outputs with period $\ell+1$, we provided conditions on the input, such that any \textit{complete} finite-length behavior of the system can be obtained.

To handle noisy and irregularly measured data, we proposed a modification of our results by introducing low rank approximation steps in the above mentioned algorithm. If successful, the algorithm returns an estimate of kernel representation of the system than can be later used to retrieve an estimate of the finite-length behavior. The accuracy of such an estimate depends on the noise levels in the data.

The results of this paper can be further extended in several directions. For instance, the guarantees provided in Section~\ref{sec_periodicSO} considered the case where the output portion of the input/output data is missing. It was observed in simulations that our proposed methods work just as well when any $p$ instances of the data sample $w(t)$ were missing, whether it is the entire output portion or a combination of input and output portions. An interesting question for future research is how to impose conditions on the available inputs (rather than assuming availability of all inputs) such that guarantees similar to those in Section~\ref{sec_periodicSO} can be shown. Studying other patterns of missing data, apart from periodically missing ones as in Section~\ref{sec_periodicSO} is also an interesting problem for future work. Furthermore, in this work we have assumed that the complexity of the system which generated the data is known. Determining a system's complexity from (noisy) input-output data is an open problem and is an interesting topic for future research. Finally, rigorous investigation of the effect of noise in the data and quantifying the uncertainty between the true and estimated finite length behaviors is an ongoing research work.

%% file: sections/appendix.tex
\appendix
\subsection{Proof of Theorem~\ref{thm_rankZ}}\label{appndx_proofThmZ}
\begin{figure*}[!t]
	\normalsize
	\begin{align}
		\bar{Z}_2 &= \begin{bmatrix}z_{[1,p],0}^{(2)} & \cdots & \bar{\zeta}_{[1,p],\ell-1}^{(2)} & \bar{\zeta}_{[1,p],\ell}^{(2)} & I_p & z_{[1,p],\ell+1}^{(2)}\\
			z_{[p+1,p(\ell+1)-n],0}^{(2)}  & \cdots & \bar{\zeta}_{[p+1,p(\ell+1)-n],\ell-1}^{(2)} & \bar{\zeta}^{(2)}_{[p+1,p(\ell+1)-n],\ell} & \underline{\zeta}^{(2)}_{[p+1,p(\ell+1)-n],\ell} & z_{[p+1,p(\ell+1)-n],\ell+1}^{(2)}\end{bmatrix}.\label{eqn_structure_of_barZ2}\\
	\mathscr{Z} \hspace{-1mm}= \hspace{-1mm}\begin{bmatrix}
		Z_1 \\ Z_2
	\end{bmatrix}\hspace{-1mm}&=\hspace{-1mm}\scalebox{0.95}{$\begingroup\setlength\arraycolsep{1.75pt}\begin{bmatrix}
		z_{[1,p(\ell+1)-n],0}^{(1)}  & \cdots & \bar{\zeta}_{[1,p(\ell+1)-n],\ell-1}^{(1)} & \underline{\zeta}^{(1)}_{[1,p(\ell+1)-n],\ell-1} & \bar{\zeta}^{(1)}_{[1,p(\ell+1)-n],\ell} & 0_{p(\ell+1)-n\times p} & z_{[1,p(\ell+1)-n],\ell+1}^{(1)}\\
		\hline
		z_{[1,p],0}^{(2)} & \cdots & \bar{\zeta}_{[1,p],\ell-1}^{(2)} & 0_{p\times p} & \bar{\zeta}_{[1,p],\ell}^{(2)} & I_p & z_{[1,p],\ell+1}^{(2)}\\
		z_{[p+1,p(\ell+1)-n],0}^{(2)}  & \cdots & \bar{\zeta}_{[p+1,p(\ell+1)-n],\ell-1}^{(2)} & 0_{p\ell-n\times p} & \bar{\zeta}^{(2)}_{[p+1,p(\ell+1)-n],\ell} & \underline{\zeta}^{(2)}_{[p+1,p(\ell+1)-n],\ell} & z_{[p+1,p(\ell+1)-n],\ell+1}^{(2)}
	\end{bmatrix}\endgroup\hspace{-1mm}.$}\label{eqn_structure_of_Z}
	\end{align}
	\hrulefill
	\vspace{-1em}
\end{figure*}
Notice that $\mathscr{Z}\in\mathbb{R}^{2p(\ell+1)-2n \times q(\ell+2)}$ has $2p(\ell+1)-2n \geq p(\ell+2)-n$ rows. Recall that if Algorithm~\ref{alg_generalalgorithm} is successful with $d=\ell+2$, the kernel representation of the system would then be given by $p(\ell+2)-n$ linearly independent rows of $\mathscr{Z}$. Therefore, our goal now is to prove that $\mathrm{rank}(\mathscr{Z})=\mathrm{dim}(\mathrm{ker}(\mathscr{H}^\top_{\ell+2}(w)))=p(\ell+2)-n$.

Consider again the matrix $\bar{H}_2$ (with the second to last block row expanded by partitioning $w$ appropriately)
\begin{equation*}
	\scalebox{0.9}{$\bar{H}_2 = \begingroup
	\setlength\arraycolsep{1.5pt}\begin{bmatrix}
		w(1) & w({\ell+2}) & \cdots & w({(n_{c}-1)(\ell+1)+1})\\
		w(2) & w({\ell+3}) & \cdots & w({(n_{c}-1)(\ell+1)+2})\\
		\vdots & \vdots & \vdots & \vdots\\
		u({\ell}) & u({2\ell+1}) & \cdots & u({(n_{c}-1)(\ell+1)+\ell})\\
		\begin{pmatrix}
			u({\ell+1})\\y(\ell+1)
		\end{pmatrix} & \begin{pmatrix}
			u({2\ell+2})\\y(2\ell+2)
		\end{pmatrix} & \cdots & \begin{pmatrix}
			u({n_{c}(\ell+1)})\\y({n_{c}(\ell+1)})
		\end{pmatrix}\\
		w(\ell+2) & w({2\ell+3}) & \cdots & w({n_{c}(\ell+1)+1})
	\end{bmatrix}\endgroup$},
\end{equation*}
and recall that rank$(\bar{H}_2)=m(\ell+2)+n$ due to Lemma~\ref{lemma_PElikeCondition}. If the block row $\begin{bmatrix}
	y(\ell+1) & y(2\ell+2) & \cdots & y({n_{c}(\ell+1)})
\end{bmatrix}$
is removed from the above matrix, we obtain
\begin{equation*}
	\scalebox{0.9}{$\tilde{H} = \begingroup
		\setlength\arraycolsep{1.5pt}\begin{bmatrix}
			w(1) & w({\ell+2}) & \cdots & w({(n_{c}-1)(\ell+1)+1})\\
			w(2) & w({\ell+3}) & \cdots & w({(n_{c}-1)(\ell+1)+2})\\
			\vdots & \vdots & \vdots & \vdots\\
			u({\ell}) & u({2\ell+1}) & \cdots & u({(n_{c}-1)(\ell+1)+\ell})\\
			u({\ell+1}) & u({2\ell+2}) & \cdots & u({(n_{c}-1)(\ell+1)+\ell+1})\\
			w(\ell+2) & w({2\ell+3}) & \cdots & w({n_{c}(\ell+1)+1})
		\end{bmatrix},\endgroup$}
\end{equation*}
where $\tilde{H}\in\mathbb{R}^{m(\ell+2)+p\ell\times n_{c}}$. Pre-multiplying by a square full rank permutation matrix $\tilde{\Pi}_r$, $\tilde{H}$ can be expressed as
\begin{align*}
	\scalebox{0.9}{$\tilde{H}$}&\scalebox{0.9}{$=\tilde{\Pi}_r\begin{bmatrix}
			\mathcal{H}_{[1,\ell+2]}(u) \\ \hline 
			\begingroup
			\setlength\arraycolsep{1.5pt}\begin{matrix}
				y(1) & y({\ell+2}) & \cdots & y({(n_{c}-1)(\ell+1)}+1)\\
				y(2) & y({\ell+3}) & \cdots & y({(n_{c}-1)(\ell+1)}+2)\\
				\vdots & \vdots & \vdots & \vdots\\
				y({\ell-1}) & y({2\ell}) & \cdots & y({(n_{c}-1)(\ell+1)+\ell-1})\\
				y({\ell+2}) & y({2\ell+3}) & \cdots & y({n_{c}(\ell+1)+1})
			\end{matrix}\endgroup
		\end{bmatrix}$}\\
	 &\scalebox{0.9}{$=\tilde{\Pi}_r \underbrace{\left[\begin{array}{c|c}
				I & 0\\ \hline
				\begin{matrix}
					D & 0 & \cdots & 0\\
					CB & D & \cdots & 0\\
					\vdots & \vdots & \vdots\\
					CA^{\ell-3}B & CA^{\ell-4}B & \cdots & 0\\
					CA^{\ell}B & CA^{\ell-1}B & \cdots & D\\
				\end{matrix} & \begin{matrix}
					C \\ CA\\ \vdots \\ CA^{\ell-2}\\ CA^{\ell+1}
				\end{matrix}
			\end{array}\right]}_{\coloneqq \mathscr{T}_3\in\mathbb{R}^{m(\ell+2)+p\ell \times m(\ell+2)+n}}\scalebox{1}{$\begin{bmatrix}
				\mathcal{H}_{[1,\ell+2]}(u)\\  X_1
			\end{bmatrix}$},$}
\end{align*}
where $X_1\coloneqq \begingroup
\setlength\arraycolsep{2pt}\begin{bmatrix}
	x(1) & x(\ell+2) & \cdots & x((n_{c}-1)(\ell+1)+1)
\end{bmatrix}\endgroup$ and the matrix $\mathscr{T}_3$ is square or has more rows than columns (since $p\ell\geq n$). Further, $\mathscr{T}_3$ has full column rank due to the identity on the upper left block and the full rank matrix on the lower right block (see Assumption~\ref{assmp_SBO}). Following the same arguments as in Lemma~\ref{lemma_PElikeCondition}, we find that $\mathrm{rank}(\tilde{H}) =m(\ell+2)+n= \mathrm{rank}(\bar{H}_2)$.

This implies that the deleted block row, i.e., $\begin{bmatrix}
	y(\ell+1) & y(2\ell+2) & \cdots & y({n_{c}(\ell+1)})
\end{bmatrix}$, can be expressed as a linear combination of the other rows of $\bar{H}_2$. In other words, we can always select $p$ rows in $\bar{Z}_2$ (without loss of generality, let them be the first $p$ rows) such that $\bar{Z}_2$ takes the form in \eqref{eqn_structure_of_barZ2} and satisfies $\bar{Z}_2\bar{H}_2=0$, where $z^{(2)}_{i,j} = \begin{bmatrix}
	\bar{\zeta}^{(2)}_{i,j} & \underline{\zeta}^{(2)}_{i,j}
\end{bmatrix}$ and $\bar{\zeta}^{(2)}_{i,j}\in\mathbb{R}^{1\times m}$ while $\underline{\zeta}^{(2)}_{i,j}\in\mathbb{R}^{1\times p}$ (the same notation is used later for $\bar{Z}_1$ in \eqref{eqn_structure_of_Z}). 

Since $\mathrm{rank}(\bar{H}_2)=\mathrm{rank}(\mathscr{H}_{\ell+2}(w))=m(\ell+2)+n$, $\bar{Z}_2$ can be extended by inserting zeros corresponding to the location of NaN entries in $H_2$ such that the matrix $\mathscr{Z}$ takes the form in \eqref{eqn_structure_of_Z} and satisfies $\mathscr{Z}\mathscr{H}_{\ell+2}(w)=0$ (see Lemma~\ref{lemma_submatrices}). Notice that $Z_1$ has full row rank, since $\bar{Z}_1$ has full row rank by definition of a basis for the left kernel of $\bar{H}_1$. Moreover, the first $p$ rows in $Z_2$ are linearly independent from the rows of $Z_1$, due to the location of the identity matrix $I_p$ in \eqref{eqn_structure_of_Z}. Therefore, it holds that $\mathrm{rank}(\mathscr{Z})=p(\ell+1)-n+p=p(\ell+2)-n$, which completes the proof.